\documentclass[a4paper,12pt]{article}
\usepackage{natbib,amsmath,amsthm,amssymb,amsfonts}
\makeatletter

\@addtoreset{equation}{section}
\makeatother

\def\ep{\varepsilon}

\def\R{\mathbb R}
\def\S{\mathbb S}
\def\N{\mathbb N}
\def\pa{\partial}
\def\b{\backslash}

\begin{document}

\title{On inverse scattering at high energies for the multidimensional Newton equation in electromagnetic field}
\author{Alexandre Jollivet}
\date{} 

\maketitle

\noindent{\bf Abstract.} We consider the multidimensional (nonrelativistic) Newton
equation in a static electromagnetic field 
$$
\ddot x = F(x,\dot x),\ F(x,\dot x)=-\nabla V(x)+B(x)\dot x,\ \dot x={dx\over dt},\ x\in C^2(\R,\R^n),
\eqno{(*)}
$$
where $V \in C^2(\R^n,\R),$ $B(x)$ is the $n\times n$ real antisymmetric matrix with elements
$B_{i,k}(x)$, $B_{i,k}\in C^1(\R^n,\R)$ (and $B$ satisfies the closure condition), and $|\pa^{j_1}_xV(x)|+|\pa^{j_2}_x
B_{i,k}(x)| \le \beta_{|j_1|}(1+|x|)^{-(\alpha+|j_1|)}$
for $x\in \R^n,$ $1\le|j_1|\le 2,$ $0\le|j_2|\le 1$, $|j_2|=|j_1|-1$, $i,k=1\ldots n$ and some $\alpha > 1$.
We give estimates and asymptotics for scattering solutions and scattering data for the equation $(*)$ for the case of small angle
scattering. We show that at high energies the velocity valued component of the scattering operator uniquely determines the X-ray 
transforms $P\nabla V$
and $PB_{i,k}$ (on sufficiently rich sets of straight lines). Applying results on inversion of the X-ray transform $P$ we obtain that for
$n\ge 2$ the velocity valued 
component of the scattering operator at high energies uniquely determines $(\nabla V,B)$. We also consider the problem of recovering
$(\nabla V,B)$ from our high energies asymptotics found for the configuration valued component of the scattering operator. Results of the
present work were obtained by developing the inverse scattering approach of [R. Novikov, 1999] for $(*)$ with $B\equiv 0$ and of
[Jollivet, 2005] for the relativistic version of $(*)$. We emphasize that there is an interesting difference in asymptotics for
scattering solutions and scattering data for $(*)$ on the one hand  and for its relativistic version on the other.
\section{Introduction}

\noindent 1.1 {\it The nonrelativistic Newton equation}

Consider the multidimensional Newton equation in an external static electromagnetic field:
\begin{eqnarray}
\ddot x(t)=F(x(t),\dot x(t))=-\nabla V(x(t))+B(x(t))\dot x(t),\label{1.1}
\end{eqnarray}
where $x(t)\in \R^n,$ $\dot x(t)={{\rm d}x\over {\rm d}t}(t),$ and $V\in C^2(\R^n,\R)$ and for any $x\in \R^n,$ $B(x)$ is a $n\times n$
antisymmetric matrix with elements $B_{i,k}(x),$  $B_{i,k}\in C^1(\R^n,\R)$, which satisfy 
\begin{equation}
{\pa B_{i,k}\over \pa x_l}(x)+{\pa B_{l,i}\over \pa x_k}(x)+{\pa B_{k,l}\over \pa x_i}(x)=0,\label{1.4B}
\end{equation}
for $x=(x_1,\ldots,x_n)\in \R^n$ and for $l,i,k=1\ldots n$. We also assume throughout this paper that $(V,B)$ satisfies the following
conditions
\begin{eqnarray}
|\pa_x^{j_1}V(x)|&\le& \beta_{|j_1|}(1+|x|)^{-\alpha-|j_1|},\ x\in \R^n,\label{1.3}\\
|\pa_x^{j_2}B_{i,k}(x)|&\le& \beta_{|j_2|+1}(1+|x|)^{-\alpha-1-|j_2|},\ x\in \R^n,\label{1.4}
\end{eqnarray}
for $|j_1|\le 2,|j_2|\le 1,$ $i,k=1\ldots n$ and some $\alpha>1$ (here $j_l$ is the multiindex $j_l=(j_{l,1},\ldots,j_{l,n})
\in (\N\cup \{0\})^n,$ $|j_l|=\sum_{k=1}^nj_{l,k}$ and $\beta_{|j_l|}$ are positive real
constants).

For equation \eqref{1.1} the energy 
\begin{equation}
E={1\over 2}|\dot x(t)|^2+V(x(t))\label{1.2}
\end{equation}
is an integral of motion.
Note that the energy $E$ does not depend on $B$ because the magnetic force $B(x)\dot x$ is orthogonal to the velocity
$\dot x$ of the particle.

For $n=3$, the equation \eqref{1.1} is the equation in $\R^n$ of motion of a nonrelativistic particle of mass $m=1$ and charge $e=1$ in an
external and static electromagnetic field described by $(V,B)$ (see, for example, Section 17 of [LL2]). 
In this equation \eqref{1.1}, $x$ denotes the position of the particle, 
$\dot x$ denotes its velocity, $\ddot x$ denotes its acceleration and $t$ denotes the time.

\vskip 4mm
\noindent 1.2 {\it Scattering data}

Under conditions \eqref{1.3}--\eqref{1.4}, the following is valid (see, for example, [S] where
classical scattering of particles in a short-range electric field is studied, 
and see [LT] where classical scattering of particles in a long-range magnetic field is studied): for any 
$(v_-,x_-)\in \R^n\times\R^n,\ v_-\neq 0,$
the equation \eqref{1.1}  has a unique solution $x\in C^2(\R,\R^n)$ such that
\begin{equation}
{x(t)=v_-t+x_-+y_-(t),}\label{1.6}
\end{equation}
where $\dot y_-(t)\to 0,\ y_-(t)\to 0,\ {\rm as}\ t\to -\infty;$  in addition for almost any 
$(v_-,x_-)\in \R^n\times \R^n,\ v_-\neq 0,$
\begin{equation}
{x(t)=v_+t+x_++y_+(t),}\label{1.7}
\end{equation}
 where $v_+\neq 0,\ v_+=a(v_-,x_-),\ x_+=b(v_-,x_-),\ \dot y_+(t)\to 0,\ \ y_+(t)\to 0,{\rm\ as\ }t \to +\infty$.

The map 
$S: (\R^n\b \{0\})\times\R^n \to (\R^n\b\{0\})\times\R^n $
given by the formulas
\begin{equation}
{v_+=a(v_-,x_-),\ x_+=b(v_-,x_-)},\label{1.8}
\end{equation}
is called the scattering map for the equation \eqref{1.1}. In addition, $a(v_-,x_-),$ $b(v_-,x_-)$ are called the scattering data for
the equation \eqref{1.1}.

By ${\cal D}(S)$ we denote the set of definition of $S$
; by ${\cal R}(S)$ we denote the range of $S$ (by definition,
if $(v_-,x_-)\in {\cal D}(S)$, then $v_-\neq 0$ and $a(v_-,x_-)\neq 0$). 

Under the conditions \eqref{1.3}--\eqref{1.4}, the map $S$  has the following simple properties:
${\cal D}(S)$  is an open set of $\R^n \times \R^n$ and ${\rm Mes}((\R^n \times \R^n) \b {\cal D}(S))=0$ for the Lebesgue
measure on $\R^n \times \R^n$ ;
the map $S:{\cal D}(S)\to {\cal R}(S)$ is continuous and preserves the element of volume ;
$a(v_-,x_-)^2=v_-^2$.
 
If $V(x)\equiv 0$ and $B(x)\equiv 0$, then $a(v_-,x_-)=v_-,\ b(v_-,x_-)=x_-,\ (v_-,x_-)\in \R^n \times \R^n,\ v_-\neq 0$. Therefore for $a(v_-,x_-),\
b(v_-,x_-)$ we will use the following representation
\begin{equation}
\begin{array}{l}
a(v_-,x_-)=v_-+a_{sc}(v_-,x_-),\\
b(v_-,x_-)=x_-+b_{sc}(v_-,x_-),
\end{array}
\hskip 1cm (v_-,x_-)\in {\cal D}(S).\label{1.9}
\end{equation}

We will use the fact that, under the conditions \eqref{1.3}--\eqref{1.4}, the map $S$ is uniquely determined by its restriction to ${\cal M}(S)={\cal
D}(S)\cap {\cal M},$ where
$$
{\cal M}=\{(v_-,x_-)\in \R^n \times \R^n|v_-\neq 0, v_-x_-=0\}.
$$
\vskip 4mm
\noindent 1.3 {\it X-ray transform}

Consider 
$$
T\S^{n-1}=\{(\theta,x)|\theta \in \S^{n-1},x\in \R^n,\theta x=0\},
$$
where $\S^{n-1}$ is the unit sphere in $\R^n$.

Consider the X-ray transform $P$ which maps each function $f$ with the properties 
$$
f\in C(\R^n,\R^m),\ |f(x)|=O(|x|^{-\beta}),\ {\rm as\ }|x|\to \infty,{\rm\ for\ some\ }\beta>1,
$$
into a function $Pf\in C(T\S^{n-1},\R^m)$ where $Pf$ is defined by
$$
Pf(\theta,x)=\int_{-\infty}^{+\infty}f(t\theta+x)dt,\ (\theta,x)\in T\S^{n-1}.
$$
Concerning the theory of the X-ray transform, the reader is referred to [R], [GGG], [Na] and [No].
\vskip 4mm
\noindent 1.4 {\it Main results of the work}

The main results of the present work consist in the small angle scattering estimates and asymptotics for the scattering data
$a_{sc}$ and $b_{sc}$ (and scattering
solutions) for the equation \eqref{1.1} and in application of these asymptotics and estimates to inverse scattering for the 
equation \eqref{1.1} at
high energies.
Our main results include, in particular, Theorem 1.1, Proposition 1.1 formulated below and Theorem 3.1 
given in Section 3.

\vskip 2mm
{\bf Theorem 1.1.} {\it
Under conditions \eqref{1.3}--\eqref{1.4}, we have
\begin{eqnarray}
\lim_{s\to+\infty} a_{sc}(s\theta,x)&=&W_{1,1}(B,\theta,x)\label{1.10a}\\
&=&\int_{-\infty}^{+\infty}B(\tau \theta+x)\theta d\tau,\nonumber
\end{eqnarray}
and
\begin{eqnarray}
&&\lim_{s\to+\infty}s\left(a_{sc}(s\theta,x)-W_{1,1}(B,\theta,x)\right)
=W_{1,2}(V,B,\theta,x)\label{1.10b}\\
&&=-P(\nabla V)(\theta,x)
+\int\limits_{-\infty}^{+\infty}\!\!B(\tau
\theta+x)(\int\limits_{-\infty}^\tau\!\!B(\sigma\theta+x)\theta d\sigma)d\tau\nonumber\\
&&+\sum_{k=1}^n\theta_k\left(\Omega_{1,1,k}(\theta,x),\ldots, \Omega_{1,n,k}(\theta,x)\right)\nonumber
\end{eqnarray}
for $(\theta,x)\in T\S^{n-1}$, $\theta=(\theta_1,\ldots,\theta_n)$, where
$$
\Omega_{1,i,k}=\int\limits_{-\infty}^{+\infty}\!\!\nabla B_{i,k}(x+\tau \theta)\circ 
\left(\int\limits_{-\infty}^\tau\!\int\limits_{-\infty}^\sigma\!\!
B(\eta\theta+x)\theta d\eta d\sigma\right)  d\tau 
$$
for $i,k=1\ldots n$ ($\circ$ denotes the usual scalar product on $\R^n$); 
in addition, we have
\begin{eqnarray}
&&\lim_{s\to+\infty} sb_{sc}(s\theta,x)=W_{2,1}(B,\theta,x)\label{1.11a}\\
&&=\int_{-\infty}^0\int_{-\infty}^\tau B(\sigma \theta+x)\theta d\sigma d\tau
-\int^{+\infty}_0\int^{+\infty}_\tau B(\sigma \theta+x)\theta d\sigma d\tau,\nonumber
\end{eqnarray}
\begin{eqnarray}
&&\lim_{s\to+\infty}s\left(sb_{sc}(s\theta,x)-W_{2,1}(B,\theta,x)\right)
=W_{2,2}(V,B,\theta,x)\label{1.11b}\\
&&=\int\limits_{-\infty}^0\!\int\limits_{-\infty}^\tau\!\!\! (-\nabla V(\sigma \theta+x))d\sigma d\tau
-\int\limits_0^{+\infty}\!\int\limits_\tau^{+\infty}\!\!\!(-\nabla V(\sigma \theta+x))d\sigma d\tau\nonumber\\
&&+\int\limits_{-\infty}^0\!\int\limits_{-\infty}^\tau\!\!\! B(\sigma \theta+x)\left(\int\limits_{-\infty}^\sigma\!\!\! B(\eta \theta+x)\theta d\eta
\right)d\sigma d\tau\nonumber\\
&&-\int\limits^{+\infty}_0\!\int\limits^{+\infty}_\tau\!\!\! B(\sigma \theta+x)\left(\int\limits_{-\infty}^\sigma\!\!\! B(\eta \theta+x)\theta 
d\eta\right)d\sigma d\tau\nonumber\\
&&+\sum_{k=1}^n\theta_k \left(\Omega_{2,1,k}(\theta,x),\ldots, \Omega_{2,n,k}(\theta,x)\right)\nonumber
\end{eqnarray}
for $(\theta,x)\in T\S^{n-1}$, $\theta=(\theta_1,\ldots,\theta_n)$, where 
\begin{eqnarray*}
\Omega_{2,i,k}(\theta,x)&=&
\int\limits_{-\infty}^0\!\int\limits_{-\infty}^\tau\!\!\!\nabla B_{i,k}(\sigma \theta+x)\circ\left(\int\limits_{-\infty}^\sigma
\!\int\limits_{-\infty}^{\eta_1}\!\!\!
B(\eta_2 \theta+x)\theta d\eta_2 d\eta_1\right)d\sigma d\tau\\
&&-\int\limits^{+\infty}_0\!\int\limits^{+\infty}_\tau\!\!\!\nabla B_{i,k}(\sigma \theta+x)\circ\left(
\int\limits_{-\infty}^\sigma\!
\int\limits_{-\infty}^{\eta_1}\!\!\!
B(\eta_2 \theta+x)\theta d\eta_2 d\eta_1\right)d\sigma d\tau
\end{eqnarray*}
for $i,k=1\ldots n$ ($\circ$ denotes the usual scalar product on $\R^n$).
}
\vskip 2mm
Theorem 1.1 gives the first two leading terms of the high energies asymptotics of the scattering data.
Theorem 1.1 follows from Theorem 3.1 (see \eqref{3.7d} and \eqref{3.7e}) formulated in Section 3.

Note that Theorem 3.1 (see \eqref{3.7d} and \eqref{3.7e}) also gives the asymptotics of  $a_{sc},$ $b_{sc}$, when the parameters  
$\alpha,$ $n,$ $s>0,$ $\theta,$ $x$ are fixed and the norm $\beta_m$ decreases to $0$ (where $\beta_m=\max(\beta_0,\beta_1,\beta_2)$),
that is  
Theorem 3.1 also gives the ``Born approximation" for the scattering data at fixed energy when
the electromagnetic field is sufficiently weak.

\vskip 2mm

{\bf Proposition 1.1.} {\it Under conditions \eqref{1.3}--\eqref{1.4}, the following statements are valid:
\begin{itemize}
\item[(i)] $W_{1,1}(B,\theta,x)$ given for all $(\theta,x)\in T\S^{n-1}$, uniquely determines $B$;

\item[(ii)] $W_{1,1}(B,\theta,x),$ $W_{1,2}(V,B,\theta,x)$ given for all $(\theta,x)\in T\S^{n-1}$, uniquely determine $(V,B)$;

\item[(iii)] if $n\ge 3$ $W_{2,1}(B,\theta,x),$ $W_{2,2}(V,B,\theta,x)$ given for all $(\theta,x)\in T\S^{n-1}$, uniquely determine $(V,B)$;

\item[(iv)] if $n=2,$ then $V$ and $B$ are not uniquely determined by
$W_{2,1}(B,\theta,x),$ $W_{2,2}(V,B,\theta,x)$ given for all $(\theta,x)\in T\S^{n-1}$.
\end{itemize}
}
\vskip 2mm

Proposition 1.1 is proved  in Section 3.  In particular, the following formula holds
 \begin{equation}
PB_{i,k}(\theta,x)=\theta_kW_{1,1}(V,B,\theta,x)_i-\theta_iW_{1,1}(V,B,\theta,x)_k\label{1.13}
\end{equation}
for $(\theta,x)\in {\cal V}_{i,k}$,  $i,k=1\ldots n,$ where ${\cal V}_{i,k}$ is the $n$-dimensional smooth manifold defined by
\begin{equation}
{\cal V}_{i,k}=\{(\theta,x)\in T\S^{n-1}|\theta_j=0,j=1\ldots n,j\neq i,j\neq k\},\label{1.12}
\end{equation}
for $i,k=1\ldots n,\ i\neq k.$ 
(To obtain \eqref{1.13} we  use the relation
$\theta_i^2+\theta_k^2=1$ for  $(\theta,x)\in {\cal V}_{i,k},$ $\theta=
(\theta_1,\ldots,\theta_n)$.)

Using \eqref{1.10a}, \eqref{1.10b}, Proposition 1.1 $(i)$  and results on inversion of the X-ray transform $P$ for $n\ge 2$ 
(see [R], [GGG], [Na], [No]) we obtain that $a_{sc}$ determines 
uniquely $\nabla V$ and $B$ at high energies. Moreover for $n\ge 2$ methods of reconstruction of $f$ from $Pf$ (see [R], [GGG], [Na], 
[No]) permit to reconstruct $\nabla V$ and $B$ from the velocity valued component $a$ of the scattering map at high energies. The 
formulas \eqref{1.11a}, \eqref{1.11b} and Proposition 1.1 show that the first two leading terms of the high energies asymptotics of $b_{sc}$ 
do not determine uniquely $(V,B)$  when $n= 2$ but that they uniquely determine $(V,B)$ when $n\ge 3$. Actually, $(V,B)$ can be
reconstructed from the first two leading terms of the high energies asymptotics of $b_{sc}$ when $n\ge 3$ 
(see the proof of Proposition 1.1 given in Section 3). 

\vskip 4mm

\noindent 1.5 {\it Historical remarks}
  
Note that inverse scattering for the classical multidimensional Newton equation at high energies was first studied by Novikov [No]
for $B\equiv 0$. Novikov proved, in particular, two formulas which link scattering data 
at high energies to the X-ray transform of $-\nabla V$ and $V$. These formulas are generalized by formulas
\eqref{1.10a}--\eqref{1.11b} of the present work for the case $B\not\equiv 0$.

Developing Novikov's approach [No], the author also studied the inverse scattering for the classical relativistic multidimensional 
Newton equation at high energies for $B\equiv 0$ [Jo1] and for $B\not\equiv 0$ [Jo2].
We emphasize that there is an interesting difference in asymptotics for
scattering solutions and scattering data for $(*)$ on the one hand  and for its relativistic version on the other.
Only the first leading term of the high energies asymptotics for the scattering data is given in [Jo1] and [Jo2]. 
In [Jo2], both $V$ and $B$ appear in this leading term.

Note also that for the classical multidimensional Newton equation in a bounded open strictly convex domain an inverse boundary value
problem at high energies was first studied in [GN]. 

Concerning the inverse scattering problem for the classical multidimensional Newton equation at fixed energy, we refer the reader to
[No], [Jo3] and references given in [No], [Jo3].

Concerning the inverse problem for \eqref{1.1} in the one-dimensional case, we can mention the works [Ab], [K], [AFC].

Concerning  the inverse scattering problem for a particle in electromagnetic field (with $B\not\equiv 0$ or $B\equiv0$) in quantim
mechanics, see references given in [Jo2].
\vskip 4mm

\noindent 1.6 {\it Structure of the paper}
  
Further, our paper is organized as follows. In Section 2 we transform the differential equation \eqref{1.1} with initial conditions 
\eqref{1.6} into a system of integral
equations which takes the form $(y_-,\dot y_-)=A_{v_-,x_-}(y_-,\dot y_-)$. Then we study $A_{v_-,x_-}$ on a suitable space and we give 
estimates for $A_{v_-,x_-}$ and for $(A_{v_-,x_-})^2$, and, in particular,  contraction estimates for $(A_{v_-,x_-})^2$ (Lemmas 2.1, 2.2, 2.3, 2.4).   
In Section 3 we give estimates and asymptotics for the deflection $y_-(t)$ from \eqref{1.6} and for  
scattering data $a_{sc}(v_-,x_-),$ $b_{sc}(v_-,x_-)$ from \eqref{1.9} (Theorem 3.1). From these estimates and asymptotics the 
four formulas \eqref{1.10a}--\eqref{1.11b} will follow when the parameters $\beta_m,$ $\alpha,$ $n,$ ${\hat v}_-,$ $x_-$ are fixed and 
$|v_-|$ increases (where $\beta_{|j|},$ $\alpha,$ $n$ are constants from \eqref{1.3}-\eqref{1.4},
$\beta_m=\max(\beta_0,\beta_1,\beta_2);$ ${\hat v}_-=v_-/|v_-|).$ In these cases $\sup|\theta(t)|$ decreases, where $\theta(t)$ denotes
the angle between the vectors ${\dot x}(t)=v_-+{\dot y}_-(t)$ and $v_-,$ and we deal with small angle scattering. 
Note that, under the conditions of Theorem 3.1, without additional assumptions, there is the estimate $\sup|\theta(t)|<
{1\over4}\pi$ and we deal with rather small angle scattering (concerning the term ``small angle scattering" see [No] and Section 20 of
[LL1]). In Section 3 we also consider the ``Born approximation" of the scattering data at fixed energy, and we prove Proposition 1.1. 
In Section 4 we prove Lemmas 2.1, 2.2, 2.3. In Section 5, we prove  Lemma 2.4. 

\vskip 4mm
\noindent{\bf Acknowledgement.} This work was fulfilled in the framework of Ph. D. thesis research under the direction of R.G. Novikov.

\section{Contraction maps}
If $x$ satisfies the differential equation \eqref{1.1} and the initial conditions \eqref{1.6}, then $x$ satisfies the system of
integral equations 
\begin{eqnarray}
x(t)&=&tv_-+x_-+\int_{-\infty}^t\int_{-\infty}^\tau F(x(s),\dot x(s))dsd\tau,\label{S2.1}\\
\dot x(t)&=&v_-+\int_{-\infty}^tF(x(s),\dot x(s))ds,\label{S2.2}
\end{eqnarray}
for $t\in \R,$ where $F(x,\dot x)=-\nabla V(x)+B(x)\dot x,$ $v_-\in \R^n\b\{0\}$.

For $y_-(t)$ of \eqref{1.6} this system takes the form
\begin{equation}
(y_-,u_-)=A_{v_-,x_-}(y_-,u_-)(t),\label{S2.3}
\end{equation}
where $u_-(t)=\dot y_-(t)$ and
\begin{eqnarray}
A_{v_-,x_-}(f,h)(t)&=&(A_{v_-,x_-}^1(f,h)(t),A_{v_-,x_-}^2(f,h)(t)),\label{18}\\
A_{v_-,x_-}^1(f,h)(t)&=&\int_{-\infty}^t A_{v_-,x_-}^2(f,h)(\tau)d\tau,\label{19}\\
A_{v_-,x_-}^2(f,h)(t)&=&\int_{-\infty}^tF(x_-+\tau v_-+f(\tau),v_-+h(\tau))d\tau,\label{20}
\end{eqnarray}
for $v_-\in \R^n\b\{0\}$.

From \eqref{S2.3}, \eqref{1.3}--\eqref{1.4} and $y_-(t)\in C^1(\R,\R^n)$, $|y_-(t)|+|\dot y_-(t)|\to 0,$ as $t\to -\infty$, it follows,
in particular, that
\begin{equation}
\begin{array}{c}
(y_-(t),\dot y_-(t))\in C(\R,\R^n)\times C(\R,\R^n)\\
\textrm{and }|\dot y_-(t)|=O(|t|^{-\alpha}),\ |y_-(t)|=O(|t|^{-\alpha+1}),\textrm{ as }t\to -\infty,
\end{array}
\label{S2.4}
\end{equation}
where $v_-\not=0$ and $x_-$ are fixed.

For nonnegative real numbers $R$ and $r$,
consider the complete metric space
\begin{eqnarray}
&&M_{T,R,r}=\{(f,h)\in C(]-\infty, T],\R^n)\times C(]-\infty, T],\R^n)\ |\nonumber\\
&& \sup_{t\in]-\infty,T]}|f(t)-th(t)|\le r,
\ \sup_{t\in]-\infty,T]}|h(t)|\le R\},\label{10}
\end{eqnarray}
with the norm $\|.\|_{\infty,T}$ defined by
\begin{equation}
\|(f,h)\|_{\infty,T}=\max\left(\sup_{t\in]-\infty,T]}|f(t)-th(t)|,\sup_{t\in]-\infty,T]}|h(t)|\right)\label{11},
\end{equation}
where $T\in ]-\infty,+\infty]$ (if $T=+\infty$, $]-\infty, T]$ must be replaced by $]-\infty,+\infty[)$).
From \eqref{S2.4} it follows that
\begin{equation}
(y_-(t),\dot y_-(t))\in M_{T,R,r}\textrm{ for some } R\textrm{ and }r \textrm{ depending on }y_-(t)\textrm{ and }T.
\label{S2.5} 
\end{equation}

\vskip 2mm

{\bf Lemma 2.1.}
{\it Let $R>0$, $0<r\le 1$ and let $(v_-,x_-)\in \R^n\times\R^n$ be such that $|v_-|> \sqrt{2}R,$ $v_-x_-=0$.
Then under conditions \eqref{1.3}--\eqref{1.4}, the following estimates are valid :
\begin{eqnarray}
\sup_{t\in]-\infty,T]}|A_{v_-,x_-}^2(f,h)(t)|&\le& \rho_{T,2}(n,\beta_1,\alpha,|v_-|,|x_-|,R)\label{2.1a}\\
&=&{2^{\alpha+1}\beta_1\sqrt{n}(1+\sqrt{n}|v_-|+\sqrt{n}R)\over
\alpha({|v_-|\over \sqrt{2}}-R)(1+{|x_-|\over \sqrt{2}}+({|v_-|\over \sqrt{2}}-R)|T|)^\alpha},\nonumber
\end{eqnarray}
\begin{equation}
\sup_{t\in]-\infty,T]}|A_{v_-,x_-}^1(f,h)(t)-tA_{v_-,x_-}^2(f,h)(t)|\le\rho_{T,1}(n,\beta_1,\alpha,|v_-|,|x_-|,R)\label{2.2a}
\end{equation}
\vskip-3mm
\begin{equation*}
={2^{\alpha+1}\beta_1\sqrt{n}(1+\sqrt{n}|v_-|+\sqrt{n}R)\over
(\alpha-1)({|v_-|\over \sqrt{2}}-R)^2(1+{|x_-|\over \sqrt{2}}+({|v_-|\over \sqrt{2}}-R)|T|)^{\alpha-1}}
,
\end{equation*}
for $T\le 0$ and $(f,h)\in M_{T,R,r}$;
\begin{eqnarray}
\sup_{t\in]-\infty,T]}|A_{v_-,x_-}^2(f,h)(t)|&\le&\rho_2(n,\beta_1,\alpha,|v_-|,|x_-|,R)\label{2.1b}\\ 
&=&{2^{\alpha+2}\beta_1\sqrt{n}(1+\sqrt{n}|v_-|+\sqrt{n}R)\over
\alpha({|v_-|\over \sqrt{2}}-R)(1+{|x_-|\over \sqrt{2}})^\alpha},\nonumber
\end{eqnarray}
\begin{equation}
\sup_{t\in]-\infty,T]}|A_{v_-,x_-}^1(f,h)(t)-tA_{v_-,x_-}^2(f,h)(t)|\le\rho_1(n,\beta_1,\alpha,|v_-|,|x_-|,R)\label{2.2b}
\end{equation}
\vskip-3mm
\begin{equation}
={2^{\alpha+2}\beta_1\sqrt{n}(1+\sqrt{n}|v_-|+\sqrt{n}R)\over
\alpha(\alpha-1)({|v_-|\over \sqrt{2}}-R)^2(1+{|x_-|\over \sqrt{2}})^{\alpha-1}}.\nonumber
\end{equation}
for $T\ge 0$ and $(f,h)\in M_{T,R,r}$.
}
\vskip 2mm
{\bf Remark 2.1.} Note that for fixed $n$, $\beta_1$, $\alpha$, $|x_-|$, $R$, we have 
\begin{eqnarray}
\rho_1(n,\beta_1,\alpha,|v_-|,|x_-|,R)\to 0, \textrm{ as }|v_-|\to +\infty;&&\label{S2.01a}\\
\rho_2(n,\beta_1,\alpha,|v_-|,|x_-|,R)\to {2^{\alpha+2}\sqrt{2}\beta_1n\over
\alpha(1+{|x_-|\over \sqrt{2}})^\alpha}, \textrm{ as }|v_-|\to +\infty&&\label{S2.01b}
\end{eqnarray}
(we used \eqref{2.1b}--\eqref{2.2b}).
\vskip 2mm
{\bf Lemma 2.2.}
{\it Let $R>0$, $0<r\le 1$ and let $(v_-,x_-)\in \R^n\times\R^n$ be such that $|v_-|> \sqrt{2}R,$ $v_-x_-=0$.
Then under conditions \eqref{1.3}--\eqref{1.4},
for $(f_1,h_1),$ $(f_2,h_2)\in M_{T,R,r}$, the following contraction estimates are valid :
\begin{eqnarray}
&&\sup_{t\in]-\infty,T]}|A_{v_-,x_-}^2(f_1,h_1)(t)-A_{v_-,x_-}^2(f_2,h_2)(t)|\le \lambda_{4,T}
\sup_{t\in]-\infty,T]}|h_1(t)-h_2(t)|\nonumber\\
&&
+\lambda_{3,T}\sup_{t\in]-\infty,T]}|f_1(t)-f_2(t)-t(h_1(t)-h_2(t))|\label{2.3a}
\end{eqnarray}
\begin{eqnarray}
&&\sup_{t\in]-\infty,T]}|\left(A_{v_-,x_-}^1(f_1,h_1)-A_{v_-,x_-}^1(f_2,h_2)\right)(t)\nonumber\\
&&-t\left(A_{v_-,x_-}^2(f_1,h_1)-A_{v_-,x_-}^2(f_2,h_2)\right)(t)|\le\lambda_{2,T}
\sup_{t\in]-\infty,T]}|h_1(t)-h_2(t)|\nonumber\\
&& 
+\lambda_{1,T}\sup_{t\in]-\infty,T]}|f_1(t)-f_2(t)-t(h_1(t)-h_2(t))|\label{2.4a},
\end{eqnarray}
for $T\le 0$, where $\lambda_{1,T},$ $\lambda_{2,T}$, $\lambda_{3,T}$ and $\lambda_{4,T}$ are given below by formulas 
\eqref{2.6a}--\eqref{2.6d};
\begin{eqnarray}
&&\sup_{t\in]-\infty,T]}|A_{v_-,x_-}^2(f_1,h_1)(t)-A_{v_-,x_-}^2(f_2,h_2)(t)|\le\lambda_4
\sup_{t\in]-\infty,T]}|h_1(t)-h_2(t)|\nonumber\\
&&
+\lambda_3\sup_{t\in]-\infty,T]}|f_1(t)-f_2(t)-t(h_1(t)-h_2(t))|\label{2.3b}
,
\end{eqnarray}
\begin{eqnarray}
&&\sup_{t\in]-\infty,T]}|\left(A_{v_-,x_-}^1(f_1,h_1)-A_{v_-,x_-}^1(f_2,h_2)\right)(t)\nonumber\\
&&-t\left(A_{v_-,x_-}^2(f_1,h_1)-A_{v_-,x_-}^2(f_2,h_2)\right)(t)|
\le\lambda_2\sup_{t\in]-\infty,T]}|h_1(t)-h_2(t)|\nonumber\\
&& 
+\lambda_1\sup_{t\in]-\infty,T]}|f_1(t)-f_2(t)-t(h_1(t)-h_2(t))|\label{2.4b},
\end{eqnarray}
for $T\ge 0$, where $\lambda_1,$ $\lambda_2$, $\lambda_3$ and $\lambda_4$ are given below by formulas \eqref{2.6e}.
}
\vskip 2mm
Lemmas 2.1, 2.2 are proved in Section 4.
\vskip 2mm
Let $R>0$, $0<r\le 1$ and let $(v_-,x_-)\in \R^n\times\R^n$ be such that $|v_-|> \sqrt{2}R.$
For $T\le0,$ real constants $\lambda_{i,T}$ for $i=1\ldots 4$, which appear in estimates \eqref{2.3a}--\eqref{2.4a} given in Lemma 2.2,
are defined by the following formulas:
\begin{equation}
\lambda_{1,T}(n,\beta_2,\alpha,|v_-|,|x_-|,R)={2^{\alpha+2}n\beta_2(1+\sqrt{n}|v_-|+\sqrt{n}R)\over
\alpha  ({|v_-|\over\sqrt{2}}-R)^2(1+{|x_-|\over \sqrt{2}}+({|v_-|\over\sqrt{2}}-R)|T|)^{\alpha}},
\label{2.6a}
\end{equation}
\begin{equation}
\lambda_{2,T}(n,\beta_1,\beta_2,\alpha,|v_-|,|x_-|,R)=2^{\alpha+1}n{\beta_1({|v_-|\over\sqrt{2}}-R)+2\beta_2(1+\sqrt{n}|v_-|+\sqrt{n}
R)
\over(\alpha -1) ({|v_-|\over\sqrt{2}}-R)^3(1+{|x_-|\over \sqrt{2}}+({|v_-|\over\sqrt{2}}-R)|T|)^{\alpha-1}
},\label{2.6b}
\end{equation}
\begin{equation}
\lambda_{3,T}(n,\beta_2,\alpha,|v_-|,|x_-|,R)={2^{\alpha+2}n\beta_2(1+\sqrt{n}|v_-|+\sqrt{n}R)\over
(\alpha+1)  ({|v_-|\over\sqrt{2}}-R)(1+{|x_-|\over \sqrt{2}}+({|v_-|\over\sqrt{2}}-R)|T|)^{\alpha+1}},\label{2.6c}
\end{equation}
\begin{equation}
\lambda_{4,T}(n,\beta_1,\beta_2,\alpha,|v_-|,|x_-|,R)=2^{\alpha+1}n{\beta_1({|v_-|\over\sqrt{2}}-R)+2\beta_2(1+\sqrt{n}|v_-|+\sqrt{n}
R)
\over\alpha  ({|v_-|\over\sqrt{2}}-R)^2(1+{|x_-|\over \sqrt{2}}+({|v_-|\over\sqrt{2}}-R)|T|)^{\alpha}},\label{2.6d}
\end{equation}
Real constants $\lambda_i$ for $i=1\ldots 4$, which appear in estimates \eqref{2.3b}-\eqref{2.4b} given in Lemma 2.2,
are defined by the following formulas:
\begin{equation}
\lambda_1={2\lambda_{1,0}\over \alpha+1},\ \lambda_2={2\lambda_{2,0}\over \alpha},\ 
\lambda_3=2\lambda_{3,0},\ \lambda_4=2\lambda_{4,0}.\label{2.6e}
\end{equation}

We define real number $\lambda_T(n,\beta_1,\beta_2,\alpha,|v_-|,|x_-|,R)$
by 
\begin{eqnarray}
\lambda_T&=&\max(\lambda_{1,T}\lambda_{3,T}+\lambda_{2,T}\lambda_{3,T}+\lambda_{3,T}\lambda_{4,T}+\lambda_{4,T}^2\label{2.8b},\\
&&\lambda_{1,T}^2+\lambda_{1,T}\lambda_{2,T}+\lambda_{2,T}\lambda_{4,T}+\lambda_{2,T}\lambda_{3,T}),\nonumber
\end{eqnarray}
for $T\le 0$ ;
we define positive real number $\lambda(n,\beta_1,\beta_2,\alpha,|v_-|,|x_-|,R)$ by
\begin{eqnarray}
\lambda&=&\max(\lambda_1\lambda_3+\lambda_2\lambda_3+\lambda_3\lambda_4+\lambda_4^2,\label{2.9b}\\
&&\lambda_1^2+\lambda_1\lambda_2+\lambda_2\lambda_4+\lambda_2\lambda_3).\nonumber
\end{eqnarray}
\vskip 2mm
{\bf Remark 2.2.} Note that for fixed $n$, $\beta_1$, $\beta_2$, $\alpha$, $|x_-|$, $R$, $T$, we have 
\begin{equation}
\begin{array}{l}
\lambda_T(n,\beta_1,\beta_2,\alpha,|v_-|,|x_-|,R)=O(|v_-|^{-1}), \textrm{ as }|v_-|\to +\infty;\\
\lambda(n,\beta_1,\beta_2,\alpha,|v_-|,|x_-|,R)=O(|v_-|^{-1}), \textrm{ as }|v_-|\to +\infty.
\end{array}\label{S2.6}
\end{equation}

\vskip 2mm
Taking into account Lemma 2.1, Lemma 2.2, we obtain the following Corollary 2.1.

\vskip 2mm
{\bf Corollary 2.1.}  
{\it Let $R>0$, $0<r\le 1$ and let $(v_-,x_-)\in \R^n\times\R^n$ be such that $|v_-|> \sqrt{2}R,$ $v_-x_-=0$.
Then under conditions \eqref{1.3}--\eqref{1.4}, the following statements are valid :
\begin{itemize}
\item[(i)] for $T\le 0$, if $\max({\rho_{T,1}\over r},{\rho_{T,2}\over R})\le 1$ then $(A_{v_-,x_-})^2$ is a map from $M_{T,R,r}$ into $M_{T,R,r}$ and $(A_{v_-,x_-})^2$ satisfies the
following inequality
\begin{equation}
\|(A_{v_-,x_-})^2(f_1,h_1)-(A_{v_-,x_-})^2(f_2,h_2)\|_{\infty,T}\le \lambda_T\|(f_1-f_2,h_1-h_2)\|_{\infty,T},\label{2.10a}
\end{equation}
for $(f_1,h_1),$ $(f_2,h_2)\in M_{T,R,r}$ ;

\item[(ii)] if $\max({\rho_1\over r},{\rho_2\over R})\le 1$ then for $T=+\infty$, 
$(A_{v_-,x_-})^2$ is a map from $M_{T,R,r}$ into $M_{T,R,r}$ and $(A_{v_-,x_-})^2$ satisfies the
following inequality
\begin{equation}
\|(A_{v_-,x_-})^2(f_1,h_1)-(A_{v_-,x_-})^2(f_2,h_2)\|_{\infty,T}\le \lambda\|(f_1-f_2,h_1-h_2)\|_{\infty,T},\label{2.10b}
\end{equation}
for $(f_1,h_1),$ $(f_2,h_2)\in M_{T,R,r}$.
\end{itemize}
(Constants $\rho_{T,1},$ $\rho_{T,2},$ $\rho_1,$ $\rho_2,$ $\lambda_T$ and $\lambda$ are respectively defined by \eqref{2.2a},
\eqref{2.1a},
\eqref{2.2b}, \eqref{2.1b}, \eqref{2.8b} and \eqref{2.9b}.) 
}

\vskip 2mm
Taking into account \eqref{S2.5} and using Lemmas 2.1, 2.2, Corollary 2.1 (see also \eqref{3.1a}--\eqref{3.1c}) and the lemma about the contraction maps we will study the
solution $(y_-(t),u_-(t))$  of the equation \eqref{S2.3} in $M_{T,R,r}$.
 
We will use also the following results (Lemmas 2.3, 2.4).
\vskip 2mm

{\bf Lemma 2.3.}
{\it  Let conditions \eqref{1.3}--\eqref{1.4} be valid. Let $R>0$, $0<r\le 1$ and let $(v_-,x_-)\in \R^n\times\R^n$ be such that 
$|v_-|> \sqrt{2}R,$ $v_- x_-=0$.
Assume $\max({\rho_2\over R},{\rho_1\over r})\le 1$ where $\rho_1,$ $\rho_2$ are respectively defined by \eqref{2.2b}, \eqref{2.1b}.
Then the following statements are valid :
\begin{equation}
[\left(A_{v_-,x_-}\right)^2]_1(f,h)(t)=k_{v_-,x_-}(f,h) t+ l_{v_-,x_-}(f,h)+H_{v_-,x_-}(f,h)(t), t\ge 0,\label{S2.10}
\end{equation}
where $\left(A_{v_-,x_-}\right)^2=([\left(A_{v_-,x_-}\right)^2]_1,[\left(A_{v_-,x_-}\right)^2]_2)$ and
\begin{equation}
k_{v_-,x_-}(f,h)=\int\limits_{-\infty}^{+\infty}\!\!F\left(x_-+sv_-+A^1_{v_-,x_-}(f,h)(s),v_-+A^2_{v_-,x_-}(f,h)(s)\right)ds,\label{2.5a}
\end{equation}
\vskip -4mm
\begin{eqnarray}
&&l_{v_-,x_-}(f,h)=\int\limits_{-\infty}^0\!\int\limits_{-\infty}^s\!\!\!F\left(x_-+\tau v_-+A^1_{v_-,x_-}(f,h)(\tau),v_-+A^2_{v_-,x_-}(f,h)(\tau)\right)d\tau ds\nonumber\\
&&-\int\limits^{+\infty}_0\!\int\limits^{+\infty}_s\!\!\!F\left(x_-+\tau v_-+A^1_{v_-,x_-}(f,h)(\tau),v_-+A^2_{v_-,x_-}(f,h)(\tau)\right)d\tau
ds\label{2.5b}
\end{eqnarray}
\begin{equation}
H_{v_-,x_-}(f,h)(t)=\int\limits_t^{+\infty}\!\int\limits_\tau^{+\infty}\!\!\!F\left(x_-+\tau
v_-+A^1_{v_-,x_-}(f,h)(\tau),v_-+A^2_{v_-,x_-}(f,h)(\tau)\right)d\tau ds,\label{2.5c}
\end{equation}
for  $t\ge 0$, $(f,h)\in M_{T,R,r}$, $T=+\infty$. In addition, the following estimates are valid :
\begin{eqnarray}
|k_{v_-,x_-}(f,h)|&\le&\rho_2(n,\beta_1,\alpha,|v_-|,|x_-|,R),\label{2.12a}\\
|l_{v_-,x_-}(f,h)|&\le&\rho_1(n,\beta_1,\alpha,|v_-|,|x_-|,R),\label{2.12b}\\
|\dot H_{v_-,x_-}(f,h)(t)|&\le&\zeta(n,\beta_1,\alpha,|v_-|,|x_-|,R,t)\label{2.12c}\\
&=&{2^{\alpha+1}\beta_1\sqrt{n}(1+\sqrt{n}|v_-|+\sqrt{n}R)\over
\alpha({|v_-|\over \sqrt{2}}-R)(1+{|x_-|\over \sqrt{2}}+({|v_-|\over \sqrt{2}}-R)t)^\alpha},\nonumber
\end{eqnarray}
\begin{eqnarray}
|H_{v_-,x_-}(f,h)(t)|&\le&\xi(n,\beta_1,\alpha,|v_-|,|x_-|,R,t)\label{2.12d}\\
&=&{2^{\alpha+1}\beta_1\sqrt{n}(1+\sqrt{n}|v_-|+\sqrt{n}R)\over\alpha
(\alpha-1)({|v_-|\over \sqrt{2}}-R)^2(1+{|x_-|\over \sqrt{2}}+({|v_-|\over \sqrt{2}}-R)t)^{\alpha-1}},\nonumber
\end{eqnarray}
for $t\ge 0$, $(f,h)\in M_{T,R,r}$, $T=+\infty$;
in addition, for $(f,h)\in M_{T,R,r}$, $T=+\infty,$ such that $(f,h)=A_{v_-,x_-}(f,h)$, we have
\begin{eqnarray}
|k_{v_-,x_-}(f,h)-k_{v_-,x_-}(0,0)|&\le&\delta_{1,1}(n,\beta_1,\beta_2,\alpha,|v_-|,|x_-|,R)\label{2.13a}\\
&=&(\lambda_2\lambda_3+\lambda_4^2)\rho_2+(\lambda_1\lambda_3+\lambda_3\lambda_4)\rho_1,\nonumber\\
|l_{v_-,x_-}(f,h)-l_{v_-,x_-}(0,0)|&\le&\delta_{2,1}(n,\beta_1,\beta_2,\alpha,|v_-|,|x_-|,R)\label{2.13b}\\
&=&(\lambda_1\lambda_2+\lambda_2\lambda_4)\rho_2+(\lambda_1^2+\lambda_2\lambda_3)\rho_1,\nonumber
\end{eqnarray}
where $\lambda_1,$ $\lambda_2,$ $\lambda_3,$ $\lambda_4$ are defined by
\eqref{2.6e}.
}
\vskip 2mm
Lemma 2.3 is proved in Section 4.
\vskip 2mm
{\bf Remark 2.3.} Note that for fixed $n$, $\beta_1$, $\beta_2$, $\alpha$, $|x_-|$, $R$, we have 
\begin{eqnarray*}
\delta_{1,1}(n,\beta_1,\beta_2,\alpha,|v_-|,|x_-|,R)&=&O(|v_-|^{-2}), \textrm{ as }|v_-|\to +\infty,\\
\delta_{2,1}(n,\beta_1,\beta_2,\alpha,|v_-|,|x_-|,R)&=&O(|v_-|^{-3}), \textrm{ as }|v_-|\to +\infty,
\end{eqnarray*}
where $\delta_{i,1},$ $i=1,2$, are defined by \eqref{2.13a}--\eqref{2.13b} (we used \eqref{2.6e} and \eqref{2.1b}--\eqref{2.2b}).
\vskip 2mm
{\bf Lemma 2.4.}
{\it  Let conditions \eqref{1.3}--\eqref{1.4} be valid. 
Let $R>0$, $0<r\le 1$ and let $(v_-,x_-)\in \R^n\times\R^n$ be such that $|v_-|> \sqrt{2}R,$ $v_- x_-=0$. 
Assume $\max({\rho_2\over R},{\rho_1\over r})\le 1$ where $\rho_1,$ $\rho_2$ are respectively defined by \eqref{2.2b}, \eqref{2.1b}.
Then the following statements are valid:
\begin{eqnarray}
&&|k_{v_-,x_-}(0,0)-w_{1,v_-,x_-}|\le\delta_{1,2}(n,\beta_1,\beta_2,\alpha,|v_-|,|x_-|,R)\label{2.14a}\\
&&={2^{\alpha+4}\sqrt{2}n^3(1+\sqrt{n}|v_-|)(2\alpha^2+\alpha-2)\beta_1(\beta_1+2\beta_2+\beta_1\beta_2)\over 
(\alpha-1)\alpha(\alpha+1){|v_-|\over \sqrt{2}}({|v_-|\over \sqrt{2}}-R)^2(1+{|x_-|\over \sqrt{2}})^{2\alpha}},\nonumber\\
&&|l_{v_-,x_-}(0,0)-w_{2,v_-,x_-}|\le\delta_{2,2}(n,\beta_1,\beta_2,\alpha,|v_-|,|x_-|,R)\label{2.14b}\\
&&={2^{\alpha+5}n^3(2\alpha+4)\beta_1(2\beta_2+\beta_1+\beta_1\beta_2)(1+\sqrt{n}|v_-|)\over
(\alpha-1)\alpha^2(\alpha+1){|v_-|\over \sqrt{2}}({|v_-|\over \sqrt{2}}-R)^3(1+{|x_-|\over \sqrt{2}})^{2\alpha-1}},\nonumber
\end{eqnarray}
where $w_{1,v_-,x_-}$ and $w_{2,v_-,x_-}$ are defined below by \eqref{2.15a} and \eqref{2.15b}.
}
\vskip 2mm
Lemma 2.4 is proved in Section 5.
\vskip 2mm
Let $(v_-,x_-)\in \R^n\times \R^n$, $v_-\not=0$. Let $\hat v_-={v_-\over |v_-|}$, $\hat v_-=({\hat v_-}^1,\ldots,{\hat v_-}^n)$.
Then vectors $w_{1,v_-,x_-}$ and $w_{2,v_-,x_-}$, which appear in \eqref{2.14a} and \eqref{2.14b}, are defined by 
\begin{eqnarray}
w_{1,v_-,x_-}&=&\int\limits_{-\infty}^{+\infty}B(\tau\hat v_-+x_-)\hat v_- d\tau-{1\over |v_-|}P(\nabla V)(
\hat v_-,x_-)d\tau
\label{2.15a}\\
&&+{1\over |v_-|}\int\limits_{-\infty}^{+\infty}\!\!\!B(\tau\hat v_-+x_-)\left(\int\limits_{-\infty}^\tau B(\sigma \hat v_-+x_-)\hat v_-
 d\sigma\right) d\tau\nonumber\\
 &&
+{1\over |v_-|}\sum_{k=1}^n{\hat v_-}^k(\Omega_{3,1,k}(v_-,x_-),\ldots,\Omega_{3,n,k}(v_-,x_-) ),\nonumber
\end{eqnarray}
where
\begin{eqnarray}
\Omega_{3,i,k}(v_-,x_-)&=&\int\limits_{-\infty}^{+\infty}\!\int\limits_0^1\!\!\nabla B_{i,k}\left(\tau\hat v_-+x_-+{\ep\over |v_-|}
\int\limits_{-\infty}^\tau\int\limits_{-\infty}^\sigma\!\! B(\eta \hat v_-
+x_-)\hat v_- d\eta d\sigma\right)\nonumber\\
&&\circ \left(\int\limits_{-\infty}^\tau\!\int\limits_{-\infty}^\sigma\!\! B(\eta\hat v_-+x_-)\hat v_-d\eta d\sigma\right) d\ep d\tau,
\end{eqnarray}
for $i,k=1\ldots n$ ($\circ$ denotes the usual scalar product on $\R^n$), and
\begin{eqnarray}
&&w_{2,v_-,x_-}={1\over|v_-|}\left(\int\limits_{-\infty}^0\!\int\limits_{-\infty}^\tau\!\!\!B(\sigma \hat v_-+x_-)\hat v_- d\sigma 
d\tau
-\int\limits_0^{+\infty}\!\int\limits_\tau^{+\infty}\!\!\!B(\sigma \hat v_-+x_-)
\hat v_- d\sigma d\tau\right)\nonumber\\
&&+{1\over |v_-|^2}\left[\int\limits_{-\infty}^0\!\int\limits_{-\infty}^\tau\!\!\! B(\sigma \hat v_-+x_-)
\left(\int\limits_{-\infty}^\sigma \!\!\!
B(\eta\hat v_-+x_-)\hat v_-d\eta \right)d\sigma d\tau\right.\label{2.15b}\\
&&\left.- \int\limits^{+\infty}_0\!\int\limits^{+\infty}_\tau\!\!\! B(\sigma \hat v_-+x_-)\left(\int\limits_{-\infty}^\sigma 
\!\!\!B(\eta\hat v_-+x_-)\hat v_-d\eta\right) d\sigma d\tau\right]\nonumber\\
&&+{1\over |v_-|^2}\sum_{k=1}^n {\hat v_-}^k(\Omega_{4,1,k}(v_-,x_-),\ldots,\Omega_{4,n,k}(v_-,x_-))\nonumber\\
&&+{1\over |v_-|^2}\left[\int\limits_{-\infty}^0\!\int\limits_{-\infty}^\tau\!\!\!(-\nabla V(\sigma \hat v_-+x_-))d\sigma d\tau-\int\limits_0^{+\infty}
\!\int\limits_\tau^{+\infty}\!\!\!(-\nabla V)(\sigma \hat v_-+x_-)
d\sigma d\tau\right],\nonumber
\end{eqnarray}
where
\begin{eqnarray}
&&\Omega_{4,i,k}(v_-,x_-)=\\
&&\int\limits_{-\infty}^0\!\int\limits_{-\infty}^\tau\!\!\! \int\limits_0^1\!\!\!\nabla B_{i,k}\left(\sigma
\hat v_-+x_-+{\ep\over |v_-|}\int\limits_{-\infty}^\sigma\!\int\limits_{-\infty}^{\eta_1}\!\!\!B(\eta_2\hat v_-+x_-)\hat v_- d\eta_2
d\eta_1\right)\nonumber\\
&& \circ \left(\int\limits_{-\infty}^\sigma \!
\int\limits_{-\infty}^{\eta_1}\!\!\! B(\eta_2\hat v_-+x_-)\hat v_- d\eta_2 d\eta_1\right) d\ep
d\sigma d\tau\nonumber\\
&&-\int\limits^{+\infty}_0\!\int\limits^{+\infty}_\tau \!\!\!\int\limits_0^1\!\!\!\nabla B_{i,k}\left(\sigma
\hat v_-+x_-+{\ep\over |v_-|}\int\limits_{-\infty}^\sigma\!\int\limits_{-\infty}^{\eta_1}\!\!\!B(\eta_2\hat v_-+x_-)\hat v_- d\eta_2
d\eta_1\right)
\nonumber\\
&&\circ \left(\int\limits_{-\infty}^\sigma 
\!\int\limits_{-\infty}^{\eta_1}\!\!\! B(\eta_2\hat v_-+x_-)\hat v_- d\eta_2 d\eta_1\right) d\ep
d\sigma d\tau\nonumber
\end{eqnarray}
for $i,k=1\ldots n$ ($\circ$ denotes the usual scalar product on $\R^n$).
\section{Small angle scattering and inverse scattering}

\noindent 3.1 {\it Small angle
scattering}

Let the constants from \eqref{1.3}--\eqref{1.4}
($\beta_{|j|},$ $\alpha$ and  $n$) and $r\in]0,1]$ be fixed, and let $r_x$ be a nonnegative real number and let $R$ be a positive 
number such that 
\begin{equation}
R>{2^{\alpha+2}\sqrt{2}\beta_1n\over
\alpha(1+{r_x\over \sqrt{2}})^\alpha}\label{S3.01}
\end{equation} 
(see \eqref{S2.01b}). Consider the real numbers  $z_1=z_1(n,\beta_1,\alpha, R,r_x)$,  $z_2=z_2(n,\beta_1,\alpha,$  $R,r, r_x)$ and 
$z_3=z_3(n,\beta_1,\beta_2,\alpha, R,r_x)$ defined as the roots of the following equations
\begin{eqnarray}
{\rho_1(n,\beta_1,\alpha,z_1,r_x,R)\over r}&=&1,\ z_1>\sqrt{2}R,\label{3.1a}\\
{\rho_2(n,\beta_1,\alpha,z_2,r_x,R)\over R}&=&1,\ z_2>\sqrt{2}R,\label{3.1b}\\
\lambda(n,\beta_1,\beta_2,\alpha,z_3,r_x,R)&=&1,\ z_3>\sqrt{2}R,\label{3.1c}
\end{eqnarray}
where $\rho_1$, $\rho_2$ and $\lambda$ are respectively defined by \eqref{2.2b}, \eqref{2.1b} and \eqref{2.9b}.

Note that from \eqref{2.2b}, \eqref{2.1b} and \eqref{2.9b} it follows that
\begin{eqnarray}
\rho_1(n,\beta_1,\alpha,s_1,r_x,R)>\rho_1(n,\beta_1,\alpha,s_2,r_x,R), \textrm{ for }\sqrt{2}R<s_1<s_2,\label{3.2a}\\
\rho_2(n,\beta_1,\alpha,s_1,r_x,R)>\rho_2(n,\beta_1,\alpha,s_2,r_x,R), \textrm{ for }\sqrt{2}R<s_1<s_2,\label{3.2b}\\
\lambda(n,\beta_1,\beta_2,\alpha,s_1,r_x,R)>\lambda(n,\beta_1,\beta_2,\alpha,s_2,r_x,R), \textrm{ for }\sqrt{2}R<s_1<s_2;\label{3.2c}
\end{eqnarray}
in addition 
\begin{eqnarray}
{\rho_2(n,\beta_1,\alpha,s,r_x,R_1)\over R_1}>{\rho_2(n,\beta_1,\alpha,s,r_x,R_2)\over R_2}, \textrm{ for } 0<R_1<R_2<{s\over
\sqrt{2}}.\label{3.3}
\end{eqnarray}

As it was already mentioned in  Introduction, under the conditions \eqref{1.3}--\eqref{1.4}, for any $(v_-,x_-)\in \R^n\times\R^n,$ $v_-\not=0,$ the equation \eqref{1.1} has a unique
solution $x\in C^2(\R,\R^n)$ with the initial conditions \eqref{1.6}. Consider the function $y_-(t)$ from \eqref{1.6}. This function
describes deflection from free motion. 
Using Corollary 2.1, the lemma about contraction maps and estimate \eqref{2.1a} of Lemma 2.1, and using Lemmas 2.3, 2.4 and the definition of $z_1$, $z_2$, $z_3$ given by
\eqref{3.1a}--\eqref{3.1c}, we obtain the following result.
\vskip 2mm

{\bf Theorem 3.1.} {\it Let conditions \eqref{1.3}--\eqref{1.4} be valid. Let $x_-\in \R^n$ and let $0<r\le 1$.
Let $R>0$  and  $v_-\in \R^n$ be such that  $R$ satisfies \eqref{S3.01} (with ``$r_x$"$=|x_-|$) and  
$|v_-|\ge \max(z_1,z_2),$ $|v_-|>z_3,$ $v_-x_-=0$, where
$z_1=z_1(n,\beta_1,\alpha, R, |x_-|)$, $z_2=z_2(n,\beta_1,\alpha, R,r,|x_-|)$ and $z_3=z_3(n,\beta_1,\beta_2,\alpha, R,|x_-|)$ are
respectively defined 
by \eqref{3.1a}, \eqref{3.1b} and \eqref{3.1c}. 
Then the deflection $y_-(t)$ has the following properties:

\begin{equation}
(y_-,\dot y_-)\in M_{T, R,r} \textrm{ for }T=+\infty ;\label{3.4}
\end{equation} 
\begin{eqnarray}
|\dot y_-(t)|&\le&{2^{\alpha+1}\beta_1\sqrt{n}(1+\sqrt{n}|v_-|+\sqrt{n}R)\over
\alpha({|v_-|\over \sqrt{2}}-R)(1+{|x_-|\over \sqrt{2}}+({|v_-|\over \sqrt{2}}-R)|t|)^\alpha},\label{3.5a}\\
|y_-(t)|&\le&{2^{\alpha+1}\beta_1\sqrt{n}(1+\sqrt{n}|v|+\sqrt{n}R)\over
\alpha(\alpha-1)({|v|\over \sqrt{2}}-R)^2(1+{|x|\over \sqrt{2}})^{\alpha-1}}, \label{3.5b} 
\end{eqnarray}
for $t\le 0$;
in addition,
\begin{equation}
y_-(t)=t a_{sc}(v_-,x_-)+b_{sc}(v_-,x_-)+h_{v_-,x_-}(t),\label{3.6}
\end{equation}
where 
\begin{eqnarray}
|h_{v_-,x_-}(t)|&\le&\xi(n,\beta_1,\alpha,|v_-|,|x_-|,R,t),\label{3.7a}\\
|\dot h_{v_-,x_-}(t)|&\le&\zeta(n,\beta_1,\alpha,|v_-|,|x_-|,R,t),\label{3.7b}\\
|a_{sc}(v_-,x_-)|&\le&\rho_2(n,\beta_1,\alpha,|v_-|,|x_-|,R),\label{3.7c}\\
|a_{sc}(v_-,x_-)-w_{1,v_-,x_-}|&\le&\delta_{1,1}+\delta_{1,2}\label{3.7d},\\
|b_{sc}(v_-,x_-)|&\le&\rho_1(n,\beta_1,\alpha,|v_-|,|x_-|,R),\\
|b_{sc}(v_-,x_-)-w_{2,v_-,x_-}|&\le&\delta_{2,1}+\delta_{2,2},\label{3.7e}
\end{eqnarray}
for $t\ge 0$, where $\xi,$ $\zeta,$ $\rho_2,$ $\rho_1$, $\delta_{1,1},$ $\delta_{1,2}$, $\delta_{2,1}$, $\delta_{2,2}$,
$w_{1,v_-,x_-}$ and $w_{2,v_-,x_-}$ are respectively defined by 
\eqref{2.12d}, \eqref{2.12c}, \eqref{2.1b}, \eqref{2.2b}, \eqref{2.13a}, \eqref{2.14a}, \eqref{2.13b}, \eqref{2.14b}, \eqref{2.15a} and
\eqref{2.15b}.
}
\vskip 2mm
Theorem 3.1 gives, in particular, estimates for the scattering process and asymptotics for the velocity valued component of the
scattering map when $\beta_1,$ $\beta_2$, $n$, $\hat v_-$, $x_-$ are fixed (where $\hat v_-={v_-\over |v_-|}$) and $|v_-|$
increases or , e.g. $\beta_1,$ $\beta_2$, $n$, $v_-$, $\hat x_-$ are fixed and $|x_-|$ increases. In these cases $\sup_{t\in \R}
|\theta(t)|$ decreases, where $\theta(t)$ denotes the angle between the vectors $\dot x(t)=v_-+\dot y_-(t)$ and $v_-$, and we deal with
small angle scattering. Note that already under the conditions of Theorem 3.1, without additional assumptions, there is the estimate
$\sup_{t\in \R}|\theta(t)|<{1\over 4}\pi$ and we deal with a rather small angle scattering.

Using Theorem 3.1 we can obtain asymptotics and estimates for small angle scattering for functions which are expressed through
$a(v_-,x_-)$ and $b(v_-,x_-)$ (e.g. see [No] for the time delay for the case $B\equiv0$).

Theorem 3.1 proves Theorem 1.1. 

\vskip 4mm

\noindent 3.2 {\it The ``Born approximation" for the scattering data at fixed energy}

The estimates \eqref{3.7d} and  \eqref{3.7e} also 
give the asymptotics of  $a_{sc},$ $b_{sc}$, when the parameters $R$, $r,$ $\alpha,$ $n,$ $|v_-|>\sqrt{2}R,$ $x_-$ are fixed and
the norm $\beta_m$ decreases to $0$ (where 
$\beta_m=\max(\beta_0,\beta_1,\beta_2)$). 
Therefore Theorem 3.1 gives also the ``Born approximation" for the scattering data at fixed energy when
the electromagnetic field is sufficiently weak. 

Let the parameters $R$, $r,$ $\alpha,$ $n,$ $s>\sqrt{2}R,$ be fixed.
Note that for fixed  $(\theta,x)\in T\S^{n-1}$, 
from \eqref{2.15a}, \eqref{2.15b},  it follows that
\begin{equation}
\tilde{w}_{i,s\theta,x}-w_{i,s\theta,x}=O(\beta_m^2),\textrm{ as }\beta_m\to 0, \textrm{ for } i=1,2.
\label{3D.1}
\end{equation}
where vectors $\tilde{w}_{1,s\theta,x}$, $\tilde{w}_{2,s\theta,x}$, are defined by 
\begin{equation}
\tilde{w}_{1,s\theta,x}=\int\limits_{-\infty}^{+\infty}B(\tau\theta+x)\theta d\tau
-{1\over s}P(\nabla V)(\theta,x),
\label{3B.1}
\end{equation}
\begin{eqnarray}
\tilde{w}_{2,s\theta,x}&=&{1\over s}\left(\int\limits_{-\infty}^0\!\int\limits_{-\infty}^\tau\!\!\!B(\sigma \theta+x)\theta d\sigma d\tau
-\int\limits_0^{+\infty}\!\int\limits_\tau^{+\infty}\!\!\!B(\sigma \theta+x)
\theta d\sigma d\tau\right)\label{3B.2}\\
&&+{1\over s^2}\left[\int\limits_{-\infty}^0\!\int\limits_{-\infty}^\tau\!\!\!(-\nabla V(\sigma \theta+x))d\sigma d\tau
-\int\limits_0^{+\infty}
\!\int\limits_\tau^{+\infty}\!\!\!(-\nabla V)(\sigma \theta+x)
d\sigma d\tau\right].\nonumber
\end{eqnarray}
From \eqref{3D.1} and \eqref{3.7d}, it follows that the leading term of the ``Born approximation" for $a_{sc}(s\theta,x)$, 
$(\theta,x)\in T\S^{n-1}$, at fixed energy, is  given by $\tilde{w}_{1,s\theta,x}$.

From \eqref{3D.1} and \eqref{3.7e}, it follows that the leading term of the ``Born approximation" for $b_{sc}(s\theta,x)$, 
$(\theta,x)\in T\S^{n-1}$, at fixed energy is  given by $\tilde{w}_{2,s\theta,x}$.

Note that 
\begin{equation}
P(\nabla V)(\theta,x)=-{s\over 2}(\tilde{w}_{1,s\theta,x}+\tilde{w}_{1,s(-\theta),x}),\label{3.2.1}
\end{equation}
\begin{equation}
\int\limits_{-\infty}^{+\infty}B(\tau\theta+x)\theta d\tau={1\over 2}(\tilde{w}_{1,s\theta,x}-\tilde{w}_{1,s(-\theta),x}),\label{3.2.2}
\end{equation}
\begin{equation}
\int\limits_{-\infty}^0\!\int\limits_{-\infty}^\tau\!\!\!B(\sigma \theta+x)\theta d\sigma d\tau
-\int\limits_0^{+\infty}\!\int\limits_\tau^{+\infty}\!\!\!B(\sigma \theta+x)
\theta d\sigma d\tau={s\over 2}(\tilde{w}_{2,s\theta,x}+\tilde{w}_{2,s(-\theta),x}),\label{3.2.3}
\end{equation}
\begin{equation}
\int\limits_{-\infty}^0\!\int\limits_{-\infty}^\tau\!\!\!(-\nabla V)(\sigma \theta+x) d\sigma d\tau
-\int\limits_0^{+\infty}\!\int\limits_\tau^{+\infty}\!\!\!(-\nabla V)(\sigma \theta+x)
 d\sigma d\tau={s^2\over 2}(\tilde{w}_{2,s\theta,x}-\tilde{w}_{2,s(-\theta),x}),\label{3.2.4}
\end{equation}
for $s>0$, $(\theta,x)\in T\S^{n-1}$.

Using \eqref{3.2.1}, \eqref{3.2.2}, \eqref{3B.1}, \eqref{1.13} and results on inversion of the X-ray transform (see [R], [GGG], [Na], 
[No]), we obtain that  for $n\ge 2$ the electromagnetic field
$(V,B)$ can be reconstructed from the leading term $\tilde{w}_{1,s\theta,x}$ of the ``Born approximation" for $a_{sc}$ at fixed energy.
We can also prove that 
$V$ for $n\ge 2$ can be reconstructed from the leading term $\tilde{w}_{2,s\theta,x}$ of the ``Born approximation" for $b_{sc}$ at fixed energy (see
\eqref{3.2.4}, \eqref{3B.2},
\eqref{3.8}). For $n\ge 3,$  $B$ can be reconstructed from the leading term $\tilde{w}_{2,s\theta,x}$ of the ``Born approximation" for $b_{sc}$ at fixed energy 
(see \eqref{3.2.3}, \eqref{3B.2} and [Jo2]). For $n=2$  the leading term $\tilde{w}_{2,s\theta,x}$ of the ``Born approximation" for $b_{sc}$ at fixed energy does not determine
uniquely $B$  (see \eqref{3.2.3} and, for example, [Jo2]).

\vskip 4mm
\noindent 3.3 {\it Proof of Proposition 1.1}

Now we prove Proposition 1.1 that deals with the reconstruction of the force field from the high energies
asymptotics we found for the scattering data.
The first item of Proposition 1.1 follows from formula \eqref{1.13} and from inversion formulas for the X-ray transform 
(see [R], [GGG], [Na], [No]). The second item follows from the first one  and from inversion formulas for the
X-ray transform. 

We prove the third item. We assume that $n\ge 3$. The magnetic field $B$ can be reconstructed from the vector $W_{2,1}(B,\theta,x)$ given for all $(\theta,x)\in T\S^{n-1}$ 
(see [Jo2]). As  $B$ is now known and $W_{2,2}(V,B,\theta,x)$ is given for all $(\theta,x)\in T\S^{n-1}$, from
\eqref{1.11b} it follows that
\begin{equation}
-PV(\theta,x)=\left(\int\limits_{-\infty}^0\!\int\limits_{-\infty}^\tau\!\!\! (-\nabla V(\sigma \theta+x))d\sigma d\tau
-\int\limits_0^{+\infty}\!\int\limits_\tau^{+\infty}\!\!\!(-\nabla V(\sigma \theta+x))d\sigma d\tau\right)\circ \theta\label{3.8}
\end{equation}
is known for all $(\theta,x)\in T\S^{n-1}$, where $\circ $ denotes the usual scalar product on $\R^n$.
Hence using also methods of reconstruction of a function from its X-ray transform (see [R], [GGG], [Na], [No]), we obtain that for $n\ge
3$, $(V,B)$ can be reconstructed from $W_{2,1}(B,\theta,x),$ $W_{2,2}(V,B,\theta,x)$ given for all $(\theta,x)\in T\S^{n-1}.$

We prove the fourth item.
Assume that $n=2$.  
We shall prove  the existence of spherical symmetric magnetic fields $B_1$ and $B_2$ satisfying \eqref{1.4} and the existence of a
spherical symmetric potential $V$ satisfying \eqref{1.3} such that $B_1\not\equiv B_2$, $V\not\equiv 0$ and 
\begin{eqnarray}
W_{2,2}(V,B_1,\theta,x)=W_{2,2}(0,B_2,\theta,x),\label{3.10a}
\end{eqnarray}
for all $(\theta,x)\in T\S^{n-1}$. Note that if $B$ is a spherical symmetric magnetic field
satisfying \eqref{1.4}, then from \eqref{1.11a} it follows that $W_{2,1}(B,\theta,x)=0$ for $(\theta,x)\in T\S^{n-1}$.

We denote by $C^\infty_0(\R^l,\R)$ the space of infinitely smooth and compactly supported function from $\R^l$ to $\R$, where $l\ge 1$.
Let $\chi \in C^\infty_0(\R,\R)$ be such that
\begin{equation}
\chi\not\equiv 0,\ 
{\rm supp}\chi\subseteq ]0,1[,\ 
\chi(x)\ge 0 \textrm{ for all }x\in \R.\label{3.11a}
\end{equation}

Consider the even functions $\tilde{f_i}\in C^\infty_0(\R,\R),$ $i=1,2$, given by the following formulas
\begin{equation}
\tilde{f_i}(q)=\chi(q)+\chi(-q)+\epsilon_i\chi(q-4)+\epsilon_i\chi(-4-q),\textrm{ for }q\in \R,\label{3.11e}
\end{equation}
where $\epsilon_1=1$ and $\epsilon_2=-1$.
Note that using \eqref{3.11a}--\eqref{3.11e} we obtain
\begin{eqnarray}
\tilde{f_1}^2\equiv\tilde{f_2}^2\label{3.11g}.
\end{eqnarray}

Using the Gelfand--Graev--Helgason range characterization of the X-ray 
transform on the Schwartz space ${\cal S}(\R^2)$  (see [GG], [H]), 
we obtain that there exists an unique function $B^i_{1,2}\in {\cal S}(\R^2)$ such that 
\begin{equation}
PB^i_{1,2}(\theta,q\theta^\bot)=\tilde{f_i}(q), \textrm{ for all } \theta\in \S^1,\ q\in \R.\label{3.12}
\end{equation} 
Note that $B^i_{1,2}\in C^\infty_0(\R^2,\R)$ since its X-ray transform is compactly supported on $T\S^2$ (see support results going
back to [C], [H] for the classical $2$-dimensional X-ray transform).

From \eqref{3.12}, it follows that for $i=1,2,$ the Fourier transform ${\cal F}B^i_{1,2}$ of the function $B^i_{1,2}$ is given by 
$$
{\cal F}B^i_{1,2}(p)=\int_{-\infty}^{+\infty}e^{-i|p|q}PB^i_{1,2}({\hat p}^\bot,q\hat
p)dq=\int_{-\infty}^{+\infty}e^{-i|p|q}\tilde{f_i}(q)dq,
$$
for $p\in \R^2,$ $p\not=0,$  ${\hat p}={p\over |p|}$ and where $\theta^\bot=(\theta_2,-\theta_1)$ for 
$\theta=(\theta_1,\theta_2)\in \S^1$.
Hence for $i=1,2,$ the Fourier transform ${\cal F}B^i_{1,2}$ is spherical symmetric. Therefore  for $i=1,2,$ $B^i_{1,2}$ is spherical
symmetric and we put
\begin{equation}
B^i_{1,2}(x)=f_i(|x|^2)\label{3.13}
\end{equation} 
for any $x\in \R^2.$
We consider the infinitely smooth and compactly supported magnetic fields $B_i$, $i=1,2,$ defined by
\begin{equation}
B_i(x)=f_i(|x|^2)
\left[\begin{matrix}
0&1\\
-1&0
\end{matrix}\right].\label{3.15}
\end{equation} 
From \eqref{3.13}, \eqref{3.12}, \eqref{3.11a}--\eqref{3.11e}, it follows that $B_1\not\equiv B_2$.
We also consider the potential $V\in C^\infty_0(\R^2,\R)$ 
defined by 
\begin{eqnarray}
&&PV(\theta,q\theta^\bot)=-\left(\int_{-\infty}^0\int_{-\infty}^\tau f_1(\sigma^2+q^2)\left(\int_{-\infty}^\sigma f_1(\eta^2
+q^2) d\eta\right) d\sigma d\tau\right.\nonumber\\
&&-\left.\int^{+\infty}_0\int^{+\infty}_\tau f_1(\sigma^2+q^2)\left(\int_{-\infty}^\sigma f_1(\eta^2
+q^2)d\eta\right) d\sigma d\tau\right)\label{3.13b}\\
&&+\int_{-\infty}^0\int_{-\infty}^\tau f_2(\sigma^2+q^2)\left(\int_{-\infty}^\sigma f_2(\eta^2
+q^2) d\eta\right) d\sigma d\tau\nonumber\\
&&-\int^{+\infty}_0\int^{+\infty}_\tau f_2(\sigma^2+q^2)\left(\int_{-\infty}^\sigma f_2(\eta^2
+q^2) d\eta\right)d\sigma d\tau\nonumber,
\end{eqnarray}
for all $\theta\in \S^1,\ q\in \R$. 

We shall prove \eqref{3.16} and \eqref{3.18}.
From \eqref{1.11b}, \eqref{3.8} and  \eqref{3.15}--\eqref{3.13b}, it follows that
\begin{equation}
W_{2,2}(V,B_1,\theta,q\theta^\bot)\circ \theta=W_{2,2}(0,B_2,\theta,q\theta^\bot)\circ \theta\label{3.16}
\end{equation}
for $q\in \R$, $\theta\in \S^1$ ($\theta=(\theta_1,\theta_2),$ $\theta^\bot=(\theta_2,-\theta_1)$).

From \eqref{3.13b}
it follows that $V$ is spherical symmetric. Hence using also \eqref{1.11b} and \eqref{3.15}  we obtain 
$$
W_{2,2}(V,B_1,\theta,q\theta^\bot)\circ \theta^\bot
=
2q\int\limits_{-\infty}^0\!\int\limits_{-\infty}^\tau\!\!\!{df_1\over ds}(s)_{|s=\sigma^2+q^2}
\left(\int\limits_{-\infty}^\sigma\!\int\limits_{-\infty}^{\eta_1}\!\!\!
f_1(\eta_2^2 +q^2) d\eta_2 d\eta_1\right)d\sigma d\tau
$$
\vskip -3mm
$$
-2q\int\limits^{+\infty}_0\!\int\limits^{+\infty}_\tau\!\!\!{df_1\over ds}(s)_{|s=\sigma^2+q^2}\left(\int\limits_{-\infty}^\sigma\!\int\limits_{-\infty}^{\eta_1}
\!\!\!f_1(\eta_2^2+q^2)d\eta_2 d\eta_1\right)d\sigma d\tau
$$
for $\theta\in \S^1,$ $q\in \R$.
Let $\theta\in \S^1$ and $q\in \R$. Integrating by parts (we remind that $f_1$ is compactly supported), we obtain
$$
W_{2,2}(V,B_1,\theta,q\theta^\bot)\circ \theta^\bot
=-2q\int\limits_{-\infty}^0\!\!\!\tau{df_1\over ds}(s)_{|s=\tau^2+q^2}
\left(\int\limits_{-\infty}^\tau\!\int\limits_{-\infty}^{\eta_1}\!\!\!
f_1(\eta_2^2 +q^2) d\eta_2 d\eta_1\right)d\tau
$$
\vskip -3mm
$$
-2q\int\limits^{+\infty}_0\!\!\!\tau{df_1\over ds}(s)_{|s=\tau^2+q^2}\left(\int\limits_{-\infty}^\tau\!\int\limits_{-\infty}^{\eta_1}
\!\!\!f_1(\eta_2^2+q^2) d\eta_2 d\eta_1\right)d\tau
$$
\vskip -3mm
$$
=q\int\limits_{-\infty}^0\!\!\!f_1(\tau^2+q^2)
\left(\int\limits_{-\infty}^\tau\!\!\!
f_1(\eta^2+q^2) d\eta \right)d\tau+q\int\limits^{+\infty}_0\!\!\! f_1(\tau^2+q^2)
\left(\int\limits_{-\infty}^\tau\!\!\!
f_1(\eta^2+q^2) d\eta\right) d\tau
$$
\vskip -3mm
\begin{equation}
=2q\left(\int\limits_0^{+\infty}\!\!\!f_1(\tau^2+q^2)d\tau\right)^2={q\over 2}\tilde{f_1}(q)^2\label{3.17}
\end{equation}
(we used the equality ${d\over d\tau}f_1(\tau^2+q^2)=2\tau{d\over d s}f_1(s)_{|s=\tau^2+q^2}$).
Using \eqref{3.11g}, \eqref{3.13} and \eqref{3.17}, we obtain 
\begin{eqnarray}
W_{2,2}(V,B_1,\theta,q\theta^\bot)\circ \theta^\bot=W_{2,2}(0,B_2,\theta,q\theta^\bot)\circ \theta^\bot.\label{3.18}
\end{eqnarray} 
Formulas \eqref{3.16} and \eqref{3.18} prove that $W_{2,2}(V,B_1,\theta,x)=W_{2,2}(0,B_2,\theta,x)$ for all $(\theta,x)\in T\S^1.$

Now it remains to prove that $V\not\equiv 0.$
Using first polar coordinates and then using \eqref{3.12}--\eqref{3.13}, we obtain that
\begin{equation}
\int_0^{+\infty}f_i(s)ds=2\int_0^{+\infty}rf_i(r^2)dr
={1\over \pi}\int_{\R^2}f_i(|x|^2)dx={1\over \pi}\int_{-\infty}^{+\infty}\tilde{f_i}(q)dq,\ i=1,2.\label{S3.1}
\end{equation}
Note that $\int_{-\infty}^{+\infty}\tilde{f_2}(q)dq=0$ and $\int_{-\infty}^{+\infty}\tilde{f_1}(q)dq=
4\int_{-\infty}^{+\infty}\chi(q)dq>0$ (we used  \eqref{3.11a}, \eqref{3.11e}). Therefore from 
\eqref{S3.1} it follows that
\begin{equation}
\left(\int_0^{+\infty}f_1(s)ds\right)^2\not=\left(\int_0^{+\infty}f_2(s)ds\right)^2.\label{3.13de}
\end{equation}
Note that for any  $q\in \R,\ i=1,2,$ 
\begin{eqnarray}
&&\int_{-\infty}^0\int_{-\infty}^\tau f_i(\sigma^2+q^2)\left(\int_{-\infty}^\sigma f_i(\eta^2
+q^2) d\eta\right) d\sigma d\tau\nonumber\\
&&-\int^{+\infty}_0\int^{+\infty}_\tau f_i(\sigma^2+q^2)\left(\int_{-\infty}^\sigma f_i(\eta^2
+q^2)d\eta\right) d\sigma d\tau\nonumber\\
&&=-\int_0^{+\infty}\int_\tau^{+\infty}f_i(\sigma^2+q^2)\left(\int_{-\sigma}^\sigma f_i(\eta^2
+q^2)d\eta\right) d\sigma d\tau.\label{3.14}
\end{eqnarray}
Assume that $V\equiv 0$, i.e. 
\begin{equation}
\int\limits_0^{+\infty}\!\int\limits_\tau^{+\infty}\!\!\!f_1(\sigma^2+q^2)\left(\int\limits_{-\sigma}^\sigma\!\!\! f_1(\eta^2
+q^2)d\eta\right) d\sigma d\tau=\int\limits_0^{+\infty}\!\int\limits_\tau^{+\infty}\!\!\!f_2(\sigma^2+q^2)\left(\int\limits_{-\sigma}^\sigma\!\!\! f_2(\eta^2
+q^2)d\eta\right) d\sigma d\tau\label{3.9}
\end{equation}
for all $q\in \R$ (we used \eqref{3.13b}, \eqref{3.14}).

For $i=1,2,$ we consider the bounded function $F_i\in C^1([0,+\infty[,\R)$ defined by 
\begin{equation}
F_i(s)=-\int^{+\infty}_sf_i(t)dt, \textrm{ for } s\in \R.\label{3.19}
\end{equation} 
Let $q\in \R.$ Note that by integrating by parts, we obtain
$$
\int\limits_0^{+\infty}\!\int\limits_\tau^{+\infty}\!\!\! f_i(\sigma^2+q^2)\left(\int\limits_{-\sigma}^\sigma\!\!\! f_i(\eta^2+q^2)d\eta\right) d\sigma d\tau\nonumber
=\int\limits_0^{+\infty}\!\!\!\tau f_i(\tau^2+q^2)\left(\int\limits_{-\tau}^\tau\!\!\! f_i(\eta^2+q^2)d\eta\right) d\tau
$$
\begin{equation}
=-\int\limits_0^{+\infty}\!\!\!F_i(\tau^2+q^2)f_i(\tau^2+q^2)d\tau\label{3.20}
\end{equation}
for $i=1,2$ (we used the equality ${d\over d\tau}F_i(\tau^2+q^2)=2\tau f_i(\tau^2+q^2)$).

From \eqref{3.20} and \eqref{3.9} and inversion of the X-ray transform (put $g_i(x)=F_i(|x|^2)f_i(|x|^2),$ $x\in \R^2$, then
$Pg_i(\theta,x)=\int_{-\infty}^{+\infty}F_i(\tau^2+x^2)f_i(\tau^2+x^2)d\tau$ for $(\theta,x)\in T\S^1$), it follows that
\begin{equation}
F_1(s)f_1(s)=F_2(s)f_2(s), \textrm{ for }s\in [0,+\infty[.
\end{equation}
Using also \eqref{3.19} ($F_i(s)\to 0$ as $s\to +\infty$) and using the equality $2F_1(s)f_1(s)={d F_1^2\over ds}(s),$ $s\in \R,$
we obtain that
$F_1^2\equiv F_2^2.$
We obtain, in particular, $F_1(0)^2=F_2(0)^2$,
which with \eqref{3.19} contradicts \eqref{3.13de}.

Proposition 1.1 is proved.\hfill$\Box$\\

{\bf Remark 3.1.} Note that there do not exist nontrivial spherical symmetric magnetic fields satisfying \eqref{1.4} (and \eqref{1.4B})
in dimension $n\ge 3$.

Note also that using \eqref{1.11b} we obtain
\begin{equation*}
W_{2,2}(V,B,\theta,x)=W_{2,2}(V,-B,\theta,x)
\end{equation*}
for $(\theta,x)\in T\S^{n-1}$ and for  $(V,B)$ satisfying \eqref{1.3}--\eqref{1.4}.

\section{Proof of Lemmas 2.1, 2.2, 2.3}
Throughout this Section, we omit index ${}_-$ for $v_-$ and $x_-$.
\vskip 4mm
\noindent 4.1 {\it Preliminary estimates}

First we prove the following Lemma.
\vskip 2mm
{\bf Lemma 4.1.}{\it
Let $(v,x)\in \R^n\times\R^n$ such that $vx=0$ and $|v|> \sqrt{2}R$.
Let $T\in ]-\infty,+\infty]$ and let $r$ be a positive real number such that $r\le  1.$
Then
\begin{eqnarray}
|f(t)|&\le& R|t|+r,\label{12}\\
|h(t)|&\le& R,\label{13}\\
1+|x+tv+f(t)|&\ge&{1\over 2}\left(1+{|x|\over \sqrt{2}}+({|v|\over \sqrt{2}}-R)|t|\right),
\label{14}\\
|v+h(t)|&\le& |v|+R,\label{14b}
\end{eqnarray}
for any $(f,h)\in {\cal M}_{T,R,r}$ and $t\le T$.
}
Under the conditions \eqref{1.3}--\eqref{1.4}, we have
\begin{eqnarray}
&&|F(x,v)|\le \beta_1\sqrt{n}(1+\sqrt{n}|v|)(1+|x|)^{-\alpha-1},\label{6}\\
&&|F(x,v)-F(x',v')|\le n\beta_1\sup_{\ep\in[0,1]}(1+|x+\ep(x'-x)|)^{-\alpha-1}|v-v'|\label{7}\\
&&+n\beta_2|x-x'|\sup_{\ep\in[0,1]}(1+|x+\ep(x'-x)|)^{-\alpha-2}(1+\sqrt{n}|v+\ep(v'-v)|),\nonumber
\end{eqnarray}
for $x,x',v,v'\in \R^n.$

\begin{proof}[Proof of Lemma 4.1]
Estimates \eqref{12} and \eqref{13} follow immediatly from \eqref{10}.
Estimate \eqref{14b} follows from \eqref{13}.
Let $(f,h)\in {\cal M}_{T,R,r}$ and $t\le T$.
As $v\circ x=0,$ we obtain
\begin{equation}
|x+tv|\ge {|x|\over \sqrt{2}}+|t|{|v|\over \sqrt{2}}.\label{16}
\end{equation}
From \eqref{12}, \eqref{16}, it follows that
\begin{eqnarray}
&&2(1+|x+tv+f(t)|)\ge 2+(1+|x+tv+f(t)|)\nonumber\\
&&\ge 2+|x+tv|-R|t|-r
\ge 2-r+{|x|\over \sqrt{2}}+|t|({|v|\over \sqrt{2}}-R).\label{15}
\end{eqnarray}
Then estimate \eqref{14} follows from \eqref{15} and the estimate $r\le 1$.

Estimates \eqref{6}-\eqref{7} follow from  conditions \eqref{1.3}-\eqref{1.4}.
\end{proof}

\noindent 4.2 {\it Proof of Lemma 2.1}

Let $(v,x)\in \R^n\times\R^n$ be fixed such that $v\circ x=0$ and $|v|> \sqrt{2}R$.
Let $r$ be a positive number such that $r\le
1$.

Let $(f,h)\in {\cal M}_{T,R,r}$.
From \eqref{20}, \eqref{6}, \eqref{14} and \eqref{14b}, it follows that
\begin{eqnarray}
&&|A_{v,x}^2(f,h)(t)|\le\beta_1\sqrt{n}\int\limits_{-\infty}^t\!\!\!
(1+\sqrt{n}|v+h(\tau)|)(1+|x+\tau v+f(\tau)|)^{-\alpha-1}d\tau\nonumber\\
&&\le2^{\alpha+1}\beta_1\sqrt{n}(1+\sqrt{n}|v|+\sqrt{n}R)
\int\limits_{-\infty}^t\!\!\!(1+{|x|\over \sqrt{2}}+({|v|\over \sqrt{2}}-R)|\tau|)^{-\alpha-1}d\tau\label{21},
\end{eqnarray}
for $t\le T$.
Hence we obtain the following estimates 
\begin{equation}
|A_{v,x}^2(f,h)(t)|\le{2^{\alpha+1}\beta_1\sqrt{n}(1+\sqrt{n}|v|+\sqrt{n}R)\over
\alpha({|v|\over \sqrt{2}}-R)(1+{|x|\over \sqrt{2}}+({|v|\over \sqrt{2}}-R)|t|)^\alpha},\label{22}
\end{equation}
for $t\le 0,$ $t\le T$ ;
\begin{equation}
|A_{v,x}^2(f,h)(t)|\le{2^{\alpha+2}\beta_1\sqrt{n}(1+\sqrt{n}|v|+\sqrt{n}R)\over
\alpha({|v|\over \sqrt{2}}-R)(1+{|x|\over \sqrt{2}})^\alpha},\label{23}
\end{equation}
for $t\ge 0,$ $t\le T$. Estimates \eqref{22}-\eqref{23} prove \eqref{2.1a} and \eqref{2.1b}.

From \eqref{19} and \eqref{22}, it follows that
\begin{eqnarray}
|t||A_{v,x}^2(f,h)(t)|&\le&{2^{\alpha+1}\beta_1\sqrt{n}(1+\sqrt{n}|v|+\sqrt{n}R)\over
\alpha({|v|\over \sqrt{2}}-R)^2(1+{|x|\over \sqrt{2}}+({|v|\over \sqrt{2}}-R)|t|)^{\alpha-1}},\label{24}\\
|A_{v,x}^1(f,h)(t)|&\le&{2^{\alpha+1}\beta_1\sqrt{n}(1+\sqrt{n}|v|+\sqrt{n}R)\over
\alpha(\alpha-1)({|v|\over \sqrt{2}}-R)^2(1+{|x|\over \sqrt{2}}+({|v|\over \sqrt{2}}-R)|t|)^{\alpha-1}},
\label{25}
\end{eqnarray}
for $t\le 0,$ $t\le T.$
Hence from \eqref{24} and \eqref{25}, it follows that
\begin{equation}
|A_{v,x}^1(f,h)(t)-tA_{v,x}^2(f,h)(t)|\le
{2^{\alpha+1}\beta_1\sqrt{n}(1+\sqrt{n}|v|+\sqrt{n}R)\over
(\alpha-1)({|v|\over \sqrt{2}}-R)^2(1+{|x|\over \sqrt{2}}+({|v|\over \sqrt{2}}-R)|t|)^{\alpha-1}}
\label{26}
\end{equation}
for $t\le 0,$ $t\le T.$

Let $t\ge 0$ and $t\le T.$
Then from \eqref{19} and  \eqref{20}, it follows that
\begin{equation}
A_{v,x}^1(f,h)(t)-tA_{v,x}^2(f,h)(t)=A_{v,x}^1(f,h)(0)-\int\limits_0^t\!\int\limits_\tau^t\!\!\!F(x+\sigma v+f(\sigma),v+h(\sigma))d\sigma d\tau.
\label{27}
\end{equation}
Using \eqref{6}, \eqref{14} and \eqref{14b}, we obtain
\begin{eqnarray}
&&\left|\int_0^t\int_\tau^tF(x+\tau v+f(\tau),v+h(\tau))d\sigma d\tau\right|\le\nonumber\\
&&\beta_1\sqrt{n}\int_0^t\int_\tau^t(1+\sqrt{n}|v+h(\sigma)|)(1+|x+\sigma v+f(\sigma)|)^{-\alpha-1}d\sigma d\tau\nonumber\\
&&\le{2^{\alpha+1}\beta_1\sqrt{n}(1+\sqrt{n}|v|+\sqrt{n}R)
\over \alpha(\alpha-1)({|v|\over \sqrt{2}}-R)^2(1+{|x|\over \sqrt{2}})^{\alpha-1}}.\label{28}
\end{eqnarray}
From \eqref{25}, it follows that
\begin{equation}
|A_{v,x}^1(f,h)(0)|\le{2^{\alpha+1}\beta_1\sqrt{n}(1+\sqrt{n}|v|+\sqrt{n}R)\over
\alpha(\alpha-1)({|v|\over \sqrt{2}}-R)^2(1+{|x|\over \sqrt{2}})^{\alpha-1}}.\label{29}
\end{equation}
From \eqref{27}--\eqref{29}, it follows that
\begin{equation}
|A_{v,x}^1(f,h)(t)-tA_{v,x}^2(f,h)(t)|\le
{2^{\alpha+2}\beta_1\sqrt{n}(1+\sqrt{n}|v|+\sqrt{n}R)\over
\alpha(\alpha-1)({|v|\over \sqrt{2}}-R)^2(1+{|x|\over \sqrt{2}})^{\alpha-1}}.
\label{30}
\end{equation}
Estimates \eqref{26}, \eqref{30} prove \eqref{2.2a}, \eqref{2.2b}.
Lemma 2.1 is proved.\hfill $\Box$
\vskip 4mm

\noindent 4.3 {\it Proof of Lemma 2.2}

Let $(f_1,h_1),$ $(f_2,h_2)\in  {\cal M}_{T,R,r}$.
From \eqref{20} and \eqref{7}, it follows that
\begin{eqnarray}
&&|A_{v,x}^2(f_1,h_1)(t)-A_{v,x}^2(f_2,h_2)(t)|\le \label{31}\\
&&n\beta_1\int\limits_{-\infty}^t\!\!\!
\sup_{\ep\in[0,1]}(1+|x+vt+\ep f_1(\tau)+(1-\ep)f_2(\tau)|)^{-\alpha-1}|h_2(\tau)-h_1(\tau)|d\tau\nonumber\\
&&+n\beta_2\int\limits_{-\infty}^t\!\!\!|f_1(\tau)-f_2(\tau)|\sup_{\ep\in[0,1]}(1+|x+vt+\ep f_1(\tau)+(1-\ep)f_2(\tau)|)^{-\alpha-2}\nonumber\\
&&\times(1+\sqrt{n}|v|+\sqrt{n}|h_1(\tau)+
\ep(h_2(\tau)-h_1(\tau))|)d\tau,\nonumber
\end{eqnarray}
for $t\le T.$
Note that
\begin{eqnarray}
|h_2(\tau)-h_1(\tau)|&\le&\sup_{\sigma\in]-\infty,T]}|h_2(\sigma)-h_1(\sigma)|,\label{32}\\
|f_2(\tau)-f_1(\tau)|&\le&\sup_{\sigma\in]-\infty,T]}|f_2(\sigma)-f_1(\sigma)-\sigma(h_1(\sigma)-h_2(\sigma))|\nonumber\\
&&+|\tau| \sup_{\sigma\in]-\infty,T]}|h_2(\sigma)-h_1(\sigma)|,\label{33}
\end{eqnarray}
for $\tau\in]-\infty,T]$.

From \eqref{31}-\eqref{33}, \eqref{14} and \eqref{14b}, it follows that
\begin{eqnarray}
&&|A_{v,x}^2(f_1,h_1)(t)-A_{v,x}^2(f_2,h_2)(t)|\nonumber\\
&&\le2^{\alpha+1}n\left[\beta_1\int\limits_{-\infty}^t\!\!\!
(1+{|x|\over \sqrt{2}}+({|v|\over\sqrt{2}}-R)|\tau|)^{-\alpha-1}d\tau+2\beta_2(1+\sqrt{n}|v|+\sqrt{n}
R)\right.\nonumber\\
&&\left.\times\int\limits_{-\infty}^t\!\!\!(1+{|x|\over \sqrt{2}}+({v\over\sqrt{2}}-R)|\tau|)^{-\alpha-2}|\tau|d\tau
\right]
\sup_{\sigma\in]-\infty,T]}|h_2(\sigma)-h_1(\sigma)|\nonumber\\
&&+2^{\alpha+2}n\beta_2(1+\sqrt{n}|v|+\sqrt{n}R)\int\limits_{-\infty}^t\!\!\!
(1+{|x|\over \sqrt{2}}+({|v|\over\sqrt{2}}-R|\tau|))^{-\alpha-2}d\tau\nonumber\\
&&\times\sup_{\sigma\in]-\infty,T]}|f_2(\sigma)-f_1(\sigma)-\sigma(h_1(\sigma)-h_2(\sigma))|\label{34}
\end{eqnarray}
(we also use the convexity of 
${\cal M}_{T,R,r}$, in order to estimate, for example, $|h_1(\tau)+
\ep(h_2(\tau)-h_1(\tau))|$ for $\tau \in]-\infty,T]$ and $\ep\in [0,1]$).
Hence we obtain the following estimates 
\begin{eqnarray}
&&|A_{v,x}^2(f_1,h_1)(t)-A_{v,x}^2(f_2,h_2)(t)|\le\nonumber\\
&&2^{\alpha+1}n{\beta_1({|v|\over\sqrt{2}}-R)+2\beta_2(1+\sqrt{n}|v|+\sqrt{n}
R)
\over\alpha  ({|v|\over\sqrt{2}}-R)^2(1+{|x|\over \sqrt{2}}+({|v|\over\sqrt{2}}-R)|t|)^{\alpha}
}
\sup_{\sigma\in]-\infty,T]}|h_2(\sigma)-h_1(\sigma)|\nonumber\\
&&+{2^{\alpha+2}n\beta_2(1+\sqrt{n}|v|+\sqrt{n}R)\over
(\alpha+1)  ({|v|\over\sqrt{2}}-R)(1+{|x|\over \sqrt{2}}+({|v|\over\sqrt{2}}-R)|t|)^{\alpha+1}}\nonumber\\
&&\times\sup_{\sigma\in]-\infty,T]}|f_2(\sigma)-f_1(\sigma)-\sigma(h_1(\sigma)-h_2(\sigma))|,\label{35}
\end{eqnarray}
for $t\le 0,$ $t \le T$ ;
\begin{eqnarray}
&&|A_{v,x}^2(f_1,h_1)(t)-A_{v,x}^2(f_2,h_2)(t)|\le\nonumber\\
&&2^{\alpha+2}n{\beta_1({|v|\over\sqrt{2}}-R)+2\beta_2(1+\sqrt{n}|v|+\sqrt{n}
R)
\over\alpha  ({|v|\over\sqrt{2}}-R)^2(1+{|x|\over \sqrt{2}})^{\alpha}
}
\sup_{\sigma\in]-\infty,T]}|h_2(\sigma)-h_1(\sigma)|\nonumber
\end{eqnarray}
\begin{eqnarray}
&&+{2^{\alpha+3}n\beta_2(1+\sqrt{n}|v|+\sqrt{n}R)\over
(\alpha+1)  ({|v|\over\sqrt{2}}-R)(1+{|x|\over \sqrt{2}})^{\alpha+1}}\nonumber\\
&&\times\sup_{\sigma\in]-\infty,T]}|f_2(\sigma)-f_1(\sigma)-\sigma(h_1(\sigma)-h_2(\sigma))|,\label{36}
\end{eqnarray}
for $t\ge 0,$ $t \le T$.
Estimates \eqref{35}, \eqref{36} prove \eqref{2.3a}, \eqref{2.3b}.

From \eqref{35}, it follows that
\begin{eqnarray}
&&|A_{v,x}^1(f_1,h_1)(t)-A_{v,x}^1(f_2,h_2)(t)|\le\nonumber\\
&&2^{\alpha+1}n{\beta_1({|v|\over\sqrt{2}}-R)+2\beta_2(1+\sqrt{n}|v|+\sqrt{n}
R)
\over\alpha (\alpha-1) ({|v|\over\sqrt{2}}-R)^3(1+{|x|\over \sqrt{2}}+({|v|\over\sqrt{2}}-R)|t|)^{\alpha-1}
}
\sup_{\sigma\in]-\infty,T]}|h_2(\sigma)-h_1(\sigma)|\nonumber\\
&&+{2^{\alpha+2}n\beta_2(1+\sqrt{n}|v|+\sqrt{n}R)\over
(\alpha+1)\alpha  ({|v|\over\sqrt{2}}-R)^2(1+{|x|\over \sqrt{2}}+({|v|\over\sqrt{2}}-R)|t|)^\alpha}\nonumber\\
&&\times
\sup_{\sigma\in]-\infty,T]}|f_2(\sigma)-f_1(\sigma)-\sigma(h_1(\sigma)-h_2(\sigma))|,\label{37}\\
&&|t||A_{v,x}^2(f_1,h_1)(t)-A_{v,x}^2(f_2,h_2)(t)|\le \nonumber\\
&&2^{\alpha+1}n{\beta_1({|v|\over\sqrt{2}}-R)+2\beta_2(1+\sqrt{n}|v|+\sqrt{n}
R)
\over\alpha  ({|v|\over\sqrt{2}}-R)^3(1+{|x|\over \sqrt{2}}+({|v|\over\sqrt{2}}-R)|t|)^{\alpha-1}
}
\sup_{\sigma\in]-\infty,T]}|h_2(\sigma)-h_1(\sigma)|\nonumber
\end{eqnarray}
\begin{eqnarray}
&&+{2^{\alpha+2}n\beta_2(1+\sqrt{n}|v|+\sqrt{n}R)\over
(\alpha+1)  ({|v|\over\sqrt{2}}-R)^2(1+{|x|\over \sqrt{2}}+({|v|\over\sqrt{2}}-R)|t|)^{\alpha}}\nonumber\\
&&\times
\sup_{\sigma\in]-\infty,T]}|f_2(\sigma)-f_1(\sigma)-\sigma(h_1(\sigma)-h_2(\sigma))|,\label{38}
\end{eqnarray}
for $t\le 0,$ $t \le T$.
Using \eqref{37}-\eqref{38}, we obtain 
\begin{eqnarray}
&|A_{v,x}^1(f_1,h_1)(t)-A_{v,x}^1(f_2,h_2)(t)-t(A_{v,x}^2(f_1,h_1)(t)-A_{v,x}^2(f_2,h_2)(t))|\le\nonumber\\
&2^{\alpha+1}n{\beta_1({|v|\over\sqrt{2}}-R)+2\beta_2(1+\sqrt{n}|v|+\sqrt{n}
R)
\over(\alpha -1) ({|v|\over\sqrt{2}}-R)^3(1+{|x|\over \sqrt{2}}+({|v|\over\sqrt{2}}-R)|t|)^{\alpha-1}
}
\sup_{\sigma\in]-\infty,T]}|h_2(\sigma)-h_1(\sigma)|\label{39}\\
&+{2^{\alpha+2}n\beta_2(1+\sqrt{n}|v|+\sqrt{n}R)\over
\alpha  ({|v|\over\sqrt{2}}-R)^2(1+{|x|\over \sqrt{2}}+({|v|\over\sqrt{2}}-R)|t|)^{\alpha}}
\sup_{\sigma\in]-\infty,T]}|f_2(\sigma)-f_1(\sigma)-\sigma(h_1(\sigma)-h_2(\sigma))|,\nonumber
\end{eqnarray}
for $t\le 0,$ $t \le T$. Estimate \eqref{2.4a} follows from \eqref{39}.

From \eqref{27}, it follows that
\begin{eqnarray}
&&|A_{v,x}^1(f_1,h_1)(t)-A_{v,x}^1(f_2,h_2)(t)-t(A_{v,x}^2(f_1,h_1)(t)-A_{v,x}^2(f_2,h_2)(t))|\nonumber\\
&&\le |A_{v,x}^1(f_1,h_1)(0)-A_{v,x}^1(f_2,h_2)(0)|\label{40}\\
&&+\int\limits_0^t\!\!\int\limits_\tau^t\!\!|F(x+\tau v+f_1(\tau),v+h_1(\tau))-F(x+\tau v+f_2(\tau),v+h_2(\tau))|d\sigma d\tau\nonumber
\end{eqnarray}
for $t\ge 0,$ $t\le T.$
Using \eqref{37}, we obtain 
\begin{eqnarray}
&&|A_{v,x}^1(f_1,h_1)(0)-A_{v,x}^1(f_2,h_2)(0)|\le{\lambda_2\over 2}\sup_{\sigma\in]-\infty,T]}|h_2(\sigma)-h_1(\sigma)|\nonumber\\
&&+{\lambda_1\over 2}
\sup_{\sigma\in]-\infty,T]}|f_2(\sigma)-f_1(\sigma)-\sigma(h_1(\sigma)-h_2(\sigma))|,\label{41}
\end{eqnarray}
where $\lambda_2=\lambda_2(n,\beta_1,\beta_2,\alpha,|v|,|x|,R)$ and $\lambda_1=\lambda_1(n,\beta_2,\alpha,|v|,|x|,R)$ are defined by 
\eqref{2.6e}.
From \eqref{7}, \eqref{14}, \eqref{14b} and \eqref{32}-\eqref{33}, it follows that
$$
\int_0^t\int_\tau^t|F(x+s v+f_1(s),v+h_1(s))-F(x+s v+f_2(s),v+h_2(s))|ds d\tau
$$
\vskip -3mm
\begin{equation}
\le{\lambda_2\over 2}
\sup_{\sigma\in]-\infty,T]}|h_2(\sigma)-h_1(\sigma)|+{\lambda_1\over 2}\sup_{\sigma\in]-\infty,T]}|f_2(\sigma)-f_1(\sigma)-\sigma(h_1(\sigma)-h_2(\sigma))|,\label{42}
\end{equation}
for $t\ge0,$ $t\le T$.

From \eqref{40}--\eqref{42}, we obtain 
$$
|A_{v,x}^1(f_1,h_1)(t)-A_{v,x}^1(f_2,h_2)(t)-t(A_{v,x}^2(f_1,h_1)(t)-A_{v,x}^2(f_2,h_2)(t))|
$$
\vskip -5mm
\begin{equation}
\le\lambda_2
\sup_{\sigma\in]-\infty,T]}|h_2(\sigma)-h_1(\sigma)|
+\lambda_1\sup_{\sigma\in]-\infty,T]}|f_2(\sigma)-f_1(\sigma)-\sigma(h_1(\sigma)-h_2(\sigma))|,\label{43}
\end{equation}
for $t\ge0,$ $t\le T$. Estimate \eqref{2.4b} follows from \eqref{43}.

Lemma 2.2 is proved.\hfill $\Box$
\vskip 4mm
\noindent 4.4 {\it Proof of Lemma 2.3}

Note that using \eqref{27} and \eqref{20} we obtain 
\begin{eqnarray}
&&A_{v,x}^1(f,h)(t)=\label{l3.1}
\int\limits_{-\infty}^0\!\int\limits^\tau_{-\infty}\!\!\!F(x+\sigma v+f(\sigma),v+h(\sigma))d\sigma d\tau\\
&&-\int\limits_0^{+\infty}\!\int\limits_\tau^{+\infty}\!\!\!F(x+\sigma v+f(\sigma),v+h(\sigma))d\sigma d\tau\nonumber\\
&&+t\int\limits_{-\infty}^{+\infty}\!\!\!F(x+\tau v+f(\tau),v+h(\tau))d\tau
+\int\limits_t^{+\infty}\int\limits_{\tau}^{+\infty}\!\!\!F(x+\sigma v+f(\sigma),v+h(\sigma))d\sigma d\tau\nonumber
\end{eqnarray}
for $t\in \R$ and $(f,h)\in M_{T,R,r},$ $T=+\infty$.

Let $T=+\infty$.
As $\max({\rho_2\over R},{\rho_1\over r})\le 1$, using Corollary 2.1 we obtain
\begin{equation}
A_{v,x}(f',h')\in M_{T,R,r},\textrm{ for any }(f',h')\in M_{T,R,r}.\label{l3.2}
\end{equation}
Let $(f,h)\in M_{T,R,r}$.
Using \eqref{l3.2} and replacing $(f,h)$ by $A_{v,x}(f,h)$ in \eqref{l3.1}, we obtain \eqref{S2.10}.
Estimates \eqref{2.12a}--\eqref{2.12d} follow from \eqref{2.5a}--\eqref{2.5c}, \eqref{l3.2}, \eqref{6}, \eqref{14} and \eqref{14b} (where we replace $(f,h)$ by $A_{v,x}(f,h)$).
Note that from \eqref{2.5a} and \eqref{2.5b} (and \eqref{1.3}--\eqref{1.4}) it follows that
\begin{eqnarray}
k_{v,x}(f',h')&=&\lim_{t\to +\infty}A_{v,x}^2(A_{v,x}(f',h'))(t),\label{l3.3a}\\
l_{v,x}(f',h')&=&\lim_{t\to +\infty}A_{v,x}^1(A_{v,x}(f',h'))(t)-tA_{v,x}^2(A_{v,x}(f',h'))(t)\label{l3.3b},
\end{eqnarray}
for any $(f',h')\in M_{T,R,r}$.

We prove \eqref{2.13a}. The proof of \eqref{2.13b} is similar to the proof of \eqref{2.13a}.
Using \eqref{l3.3a}, \eqref{l3.2} and applying  \eqref{2.3b} (``$(f_1,h_1)=A_{v,x}(f,h)$" and ``$(f_2,h_2)=A_{v,x}(0,0)$"), we obtain
\begin{eqnarray}
&&|k_{v,x}(f,h)-k_{v,x}(0,0)|\le \lambda_4
\sup_{t\in]-\infty,+\infty[}|A_{v_-,x_-}^2(f,h)(t)-A_{v_-,x_-}^2(0,0)(t)|\nonumber\\
&&
+\lambda_3\sup_{t\in]-\infty,+\infty[}|\left(A_{v_-,x_-}^1(f,h)-A_{v_-,x_-}^1(0,0)\right)(t)\nonumber\\
&&-t\left(A_{v_-,x_-}^2(f,h)-A_{v_-,x_-}^2(0,0)\right)(t)|.
\label{l3.4}
\end{eqnarray}
Using \eqref{l3.4}, \eqref{2.3b}--\eqref{2.4b}, we obtain
\begin{eqnarray}
&&|k_{v_-,x_-}(f,h)-k_{v_-,x_-}(0,0)|\le
(\lambda_2\lambda_3+\lambda_4^2)\sup_{t\in]-\infty,+\infty[}|h(t)|\nonumber\\
&&+
(\lambda_1\lambda_3+\lambda_3\lambda_4)\sup_{t\in]-\infty,+\infty[}|f(t)-th(t)|.\label{l3.5}
\end{eqnarray}
Assume that $(f,h)=A_{v,x}(f,h)$. Then from \eqref{2.1b}--\eqref{2.2b}, it follows that
$$
|h(t)|\le \rho_2\textrm{ and }|f(t)-th(t)|\le \rho_1 \textrm{ for }t\in \R. 
$$
These two latter estimates with \eqref{l3.5} prove \eqref{2.13a}.
\hfill $\Box$


\section{Proof of Lemma 2.4}

Throughout this Section, we omit index ${}_-$ for $v_-$ and $x_-$.

We shall prove \eqref{2.14a}. 
Note that using changes of variables and the equality $B_{i,k}(x+\sigma v+\omega)=B_{i,k}(x+\sigma v)+\int_0^1\nabla B_{i,k}(x+\sigma
v+\ep\omega)\circ\omega d\ep$ for $\sigma\in \R,$ $\omega\in \R^n$ (where $\circ$ denotes the usual scalar product on $\R^n$), 
we obtain

\begin{eqnarray*}
w_{1,v,x}&=&-\int\limits_{-\infty}^{+\infty}\!\!\!\nabla V(sv+x)ds
+\int\limits_{-\infty}^{+\infty}\!\!\!B(sv+x)\left(\int\limits_{-\infty}^sB(\tau v+x)v d\tau\right) ds\\
&&
+\int\limits_{-\infty}^{+\infty}\!\!\!B\left(sv+x+
\int\limits_{-\infty}^s\int\limits_{-\infty}^\tau B(\sigma v+x)vd\sigma d\tau\right)v ds.
\end{eqnarray*}
Therefore, from \eqref{2.5a} and \eqref{20} (we remind that $F(x,v)=-\nabla V(x)+B(x)v$) it follows that
\begin{equation}
|k_{v,x}(0,0)-w_{1,v,x}|\le\sum_{i=1}^4\Delta_{1,i},\label{l4.1}
\end{equation}
where
\begin{equation}
\Delta_{1,1}=\int\limits_{-\infty}^{+\infty}\!\!\!\left|\nabla V(sv+x+A_{v,x}^1(0,0)(s))-\nabla V(sv+x)\right|ds,\label{l4.2a}
\end{equation}
\begin{equation}
\Delta_{1,2}=\int\limits_{-\infty}^{+\infty}\left|B(sv+x+A^1_{v,x}(0,0)(s))\left(\int\limits_{-\infty}^s\!\!\!\nabla V(\tau
v+x)d\tau\right)\right| ds,
\label{l4.2b}
\end{equation}
\begin{equation}
\Delta_{1,3}=\int\limits_{-\infty}^{+\infty}\left|\left(B(sv+x+A_{v,x}^1(0,0)(s))-B(sv+x)\right)\left(\int\limits_{-\infty}^s\!\!\!B(\tau
v+x)vd\tau\right)\right| ds,\label{l4.2c}
\end{equation}
\begin{eqnarray}
\Delta_{1,4}&=&\int\limits_{-\infty}^{+\infty}\left|\left(B(sv+x+A_{v,x}^1(0,0)(s))
\phantom{\int\limits_{-\infty}^s\!\int\limits_{-\infty}^\tau}\right.\right.\label{l4.2d}\\
&&\left.\left.-B\left(
sv+x+\int\limits_{-\infty}^s\!\int\limits_{-\infty}^\tau\!\!\! B(\sigma v+x)vd\sigma d\tau\right)
\right)v\right| ds.\nonumber
\end{eqnarray}

We shall estimate each $\Delta_{1,i}$, $i=1\ldots 4$.
First note that using Corollary 2.1 and the inequality $\max({\rho_2\over R},{\rho_1\over r})\le 1$ we obtain, in particular, 
\begin{equation}
A_{v,x}(0,0)\in M_{T,R,r},\ T=+\infty.\label{l4.3}
\end{equation}
Note also that using \eqref{1.3}--\eqref{1.4} and the estimate $|x+\sigma v|\ge{|x|\over \sqrt{2}}+|\sigma|{|v|\over \sqrt{2}},$ 
$\sigma \in \R$ (we remind that $x\circ v=0$), we obtain

\begin{equation}
|\nabla V(\sigma v+x)|\le \beta_1\sqrt{n}(1+{|x|\over \sqrt{2}}+|\sigma|{|v|\over \sqrt{2}})^{-\alpha-1},\label{l4.S1}
\end{equation}

\begin{equation}
|B(\sigma v+x)v|\le \beta_1n|v|(1+{|x|\over \sqrt{2}}+|\sigma|{|v|\over \sqrt{2}})^{-\alpha-1},\label{l4.S2}
\end{equation}
for $\sigma \in \R$.

We remind that 
\begin{eqnarray}
A_{v,x}^1(0,0)(s)&=&\int_{-\infty}^s\int_{-\infty}^\tau (-\nabla V)(\sigma v+s)d\sigma d\tau\nonumber\\
&&+\int_{-\infty}^s\int_{-\infty}^\tau B(\sigma v+x)v d\sigma d\tau,\label{l4.9}
\end{eqnarray}
for $s\in \R$.

We shall use the following estimate \eqref{l4.4}: from \eqref{l4.S1}-\eqref{l4.9}, it follows that 
\begin{eqnarray}
|A_{v,x}^1(0,0)(s)|\le \beta_1\sqrt{n}(1+\sqrt{n}|v|)\int\limits_{-\infty}^s\!\int\limits_{-\infty}^\tau\!\!\!
(1+{|x|\over \sqrt{2}}+|\sigma|{|v|\over \sqrt{2}})^{-\alpha-1}d\sigma d\tau\nonumber&&\\
\le \beta_1\sqrt{n}(1+\sqrt{n}|v|)\left(\int\limits_{-\infty}^0\!\int\limits_{-\infty}^\tau\!\!\!
(1+{|x|\over \sqrt{2}}+|\sigma|{|v|\over \sqrt{2}})^{-\alpha-1}d\sigma d\tau\right.\nonumber&&\\
\left.+|s|\int\limits_{-\infty}^{+\infty}\!\!\!
(1+{|x|\over \sqrt{2}}+|\tau|{|v|\over \sqrt{2}})^{-\alpha-1}d\tau \right)\nonumber&&\\
\le{2\beta_1\sqrt{n}(1+\sqrt{n}|v|)\over \alpha(\alpha-1)|v|^2(1+{|x|\over \sqrt{2}})^{\alpha-1}}
+|s|{2\sqrt{2}\beta_1\sqrt{n}(1+\sqrt{n}|v|)\over \alpha|v|(1+{|x|\over \sqrt{2}})^{\alpha}},&&
\label{l4.4}
\end{eqnarray}
for $s\in \R$.

Using \eqref{1.3}, \eqref{l4.3} and \eqref{14}, we obtain 
\begin{eqnarray}
&&\left|\nabla V(\sigma v+x+A_{v,x}^1(0,0)(\sigma))-\nabla V(\sigma v+x)\right|\nonumber\\
&&\le 2^{\alpha+2}\beta_2n (1+{|x|\over\sqrt{2}}+({|v|\over \sqrt{2}}-R)|\sigma|)^{-\alpha-2}|A_{v,x}^1(0,0)(\sigma)|\label{l4.5},
\end{eqnarray}
for all $\sigma \in \R$.

From \eqref{l4.2a}, \eqref{l4.4}, \eqref{l4.5}, it follows that
\begin{eqnarray}
\Delta_{1,1}
&\le& n2^{\alpha+2}\beta_2\int_{-\infty}^{+\infty}(1+{|x|\over\sqrt{2}}+({|v|\over
\sqrt{2}}-R)|s|)^{-\alpha-2}|A_{v,x}^1(0,0)(s)|ds\nonumber \\
&\le& {2^{\alpha+3}n^{3/2}(2\alpha^2+\alpha-2)\beta_1\beta_2(1+\sqrt{n}|v|)\over (\alpha-1)\alpha^2(\alpha+1){|v|\over\sqrt{2}}
({|v|\over \sqrt{2}}-R)^2
(1+{|x|\over \sqrt{2}})^{2\alpha}}.\label{l4.6}
\end{eqnarray}

Similarly, by using \eqref{l4.2c} (and by using \eqref{1.4} instead of \eqref{1.3}) and \eqref{l4.S2} we obtain
\begin{eqnarray}
\Delta_{1,3}&
\le& 2^{\alpha+2}\beta_1\beta_2n^{5\over 2}\int_{-\infty}^{+\infty}(1+({|v|\over\sqrt{2}}-R)|s|+{|x|\over\sqrt{2}})^{-\alpha-2}
|A_{v,x}^1(0,0)(s)|\nonumber\\
&&\times\left(\int_{-\infty}^{s}
(1+{|v|\over\sqrt{2}}|\tau|+{|x|\over\sqrt{2}})^{-\alpha-1}|v|d\tau \right)ds\nonumber\\
&\le& {2^{\alpha+4}\sqrt{2}(2\alpha^2+\alpha-2)n^3\beta_1^2\beta_2(1+\sqrt{n}|v|)\over
(\alpha-1)\alpha^3(\alpha+1){|v|\over \sqrt{2}}({|v|\over\sqrt{2}}-R)^2(1+{|x|\over\sqrt{2}})^{3\alpha}}.\label{l4.7}
\end{eqnarray}

Using \eqref{l4.2b}, \eqref{1.3}-\eqref{1.4}, \eqref{l4.3} and \eqref{14}, we obtain
\begin{eqnarray*}
\Delta_{1,2}
&\le&\beta_1^2n^{3/2}2^{\alpha+1}\int_{-\infty}^{+\infty}(1+{|x|\over \sqrt{2}}+({|v|\over \sqrt{2}}-R)|s|)^{-\alpha-1}\nonumber\\
&&\times\left(\int_{-\infty}^s(1+{|x|\over \sqrt{2}}+{|v|\over \sqrt{2}}|\tau|)^{-\alpha-1}d\tau\right) ds\nonumber\\
&\le &{n^{3/2}2^{\alpha+3}\beta_1^2\over \alpha^2{|v|\over \sqrt{2}}({|v|\over \sqrt{2}}-R)(1+{|x|\over \sqrt{2}})^{2\alpha}}.\label{l4.8}
\end{eqnarray*}

Using \eqref{l4.2d}, \eqref{l4.9}, growth property of the elements of $B$ \eqref{1.4}, and using \eqref{1.3} and the assumption 
$\max({\rho_1\over r},{\rho_2\over R})\le 1$ and \eqref{14}, we obtain
\begin{eqnarray}
\Delta_{1,4}
&\le& |v|n^{3/2}2^{\alpha+2}\beta_2\int_{-\infty}^{+\infty}(1+{|x|\over\sqrt{2}}+({|v|\over \sqrt{2}}-R)|s|)^{-\alpha-2}\nonumber\\
&&\times
\left(\int_{-\infty}^s\int_{-\infty}^\tau \left|\nabla V(x+\sigma v)\right|d\sigma d\tau\right) ds\nonumber\\
&\le&|v|n 2^{\alpha+2}\beta_1\beta_2\int_{-\infty}^{+\infty}(1+{|x|\over\sqrt{2}}+({|v|\over \sqrt{2}}-R)|s|)^{-\alpha-2}\nonumber\\
&&\times
\left(\int_{-\infty}^s\int_{-\infty}^\tau(1+{|x|\over\sqrt{2}}+{|v|\over \sqrt{2}}|\sigma|)^{-\alpha-1}d\sigma d\tau\right) ds\nonumber\\
&\le& {n^{2}2^{\alpha+3}\sqrt{2}(2\alpha^2+\alpha-2)
\beta_1\beta_2\over \alpha^2(\alpha+1)(\alpha-1)({|v|\over \sqrt{2}}-R)^2(1+{|x|\over \sqrt{2}})^{2\alpha}}.\label{l4.10}
\end{eqnarray}
Estimate \eqref{2.14a} follows from \eqref{l4.1} and \eqref{l4.6}-\eqref{l4.10}.

We shall prove \eqref{2.14b}. 
Note that using changes of variables and the equality $B_{i,k}(x+\sigma v+\omega)=B_{i,k}(x+\sigma v)+\int_0^1\nabla B_{i,k}(x+\sigma
v+\ep\omega)\circ\omega d\ep$ for $\sigma\in \R,$ $\omega\in \R^n$ (where $\circ$ denotes the usual scalar product on $\R^n$), we obtain

\begin{eqnarray*}
w_{2,v,x}&=&\int\limits_{-\infty}^0\!\int\limits_{-\infty}^s\!\!\!B\left(\sigma v+x+
\int\limits_{-\infty}^\sigma\!\int\limits_{-\infty}^{\eta_1}\!\!\!B(\eta_2v+x)v d\eta_2
d\eta_1
\right) v d\sigma ds
\\
&&
-\int\limits_0^{+\infty}\!\int\limits_s^{+\infty}\!\!\!B\left(\sigma  v+x+
\int\limits_{-\infty}^\sigma\!\int\limits_{-\infty}^{\eta_1}\!\!\!B(\eta_2v+x)v d\eta_2
d\eta_1\right)
 v d\sigma ds\\
&&+\int\limits_{-\infty}^0\!\int\limits_{-\infty}^\tau\!\!\! B(\sigma v+x)\left(\int\limits_{-\infty}^\sigma \!\!\!
B(\eta v+x) vd\eta \right)d\sigma d\tau\\
&&- \int\limits^{+\infty}_0\!\int\limits^{+\infty}_\tau\!\!\! B(\sigma v+x)\left(\int\limits_{-\infty}^\sigma 
\!\!\!B(\eta v+x) vd\eta \right) d\sigma d\tau\\
&&+\int\limits_{-\infty}^0\!\int\limits_{-\infty}^s\!\!\!(-\nabla V(\sigma v+x))d\sigma ds-\int\limits_0^{+\infty}
\!\int\limits_s^{+\infty}\!\!\!(-\nabla V)(\sigma v+x)
d\sigma ds,
\end{eqnarray*}
Therefore, from \eqref{2.5b} and \eqref{20} (we remind that $F(x,v)=-\nabla V(x)+B(x)v$) it follows that
\begin{equation}
|l_{v,x}(0,0)-w_{2,v,x}|\le\sum_{i=1}^6\Delta_{2,i},\label{l4.11}
\end{equation}
where
\begin{equation}
\Delta_{2,1}=\int\limits_{-\infty}^0\!\int\limits_{-\infty}^s\!\!\!\left|\nabla V(\sigma v+x+A_{v,x}^1(0,0)(\sigma))
-\nabla V(\sigma v+x)\right|d\sigma ds,\label{l4.12a}
\end{equation}
\begin{equation}
\Delta_{2,2}=\int\limits_0^{+\infty}\!\int\limits_s^{+\infty}\!\!\!\left|\nabla V(\sigma v+x+A_{v,x}^1(0,0)(\sigma))
-\nabla V(\sigma v+x)\right|
d\sigma ds,\label{l4.12b}
\end{equation} 
\begin{eqnarray}
\Delta_{2,3}&=&
\left|\int\limits_{-\infty}^0\!\int\limits_{-\infty}^\tau\!\!\! B(\sigma v+x+A_{v,x}^1(0,0)(\sigma))\left(\int\limits_{-\infty}^\sigma\!\!\! \nabla
V(\eta v+x)d\eta\right) d\sigma d\tau \right.\label{l4.12c}\\
&&- \left.\int\limits_0^{+\infty}\!\int\limits_\tau^{+\infty}\!\!\! B(\sigma v+x+A_{v,x}^1(0,0)(\sigma))\left(\int\limits_{-\infty}^\sigma\!\!\! \nabla
V(\eta v+x)d\eta\right) d\sigma d\tau \right|,\nonumber
\end{eqnarray}
\begin{equation}
\Delta_{2,4}=
\left|\int\limits_{-\infty}^0\!\int\limits_{-\infty}^\tau\!\!\! \left(B(\sigma v+x+A_{v,x}^1(0,0)(\sigma))
-B(\sigma v+x)\right)\left(\int\limits_{-\infty}^\sigma\!\!\! B(\eta v+x)v d\eta\right) d\sigma d\tau \right.\label{l4.12d}
\end{equation}
$$
- \left.\int\limits_0^{+\infty}\!\int\limits_\tau^{+\infty}\!\!\! \left(B(\sigma v+x+A_{v,x}^1(0,0)(\sigma))
-B(\sigma v+x)\right)\left(\int\limits_{-\infty}^\sigma\!\!\!
B(\eta v+x)v d\eta\right) d\sigma d\tau \right|,\nonumber
$$
\begin{eqnarray}
\Delta_{2,5}&=&
\left|\int\limits_{-\infty}^0\!\int\limits_{-\infty}^\tau\!\!\! \left(B(\sigma v+x+A_{v,x}^1(0,0)(\sigma))
\phantom{\int\limits_{-\infty}^\sigma\!\int\limits_{-\infty}^{\eta_1}}\right.\right.\label{l4.12e}\\
&&\left.\left.-B\left(\sigma v+x+\int\limits_{-\infty}^\sigma\!\int\limits_{-\infty}^{\eta_1}B(\eta_2v+x)v d\eta_2d\eta_1\right)
\right)v
 d\sigma d\tau \right|,\nonumber
\end{eqnarray}
\begin{eqnarray}
\Delta_{2,6}&=&
\left|\int\limits^{+\infty}_0\!\int\limits^{+\infty}_\tau\!\!\! \left(B(\sigma v+x+A_{v,x}^1(0,0)(\sigma))
\phantom{\int\limits_{-\infty}^\sigma\!\int\limits_{-\infty}^{\eta_1}}\right.\right.\label{l4.12f}\\
&&\left.\left.-B\left(\sigma v+x+\int\limits_{-\infty}^\sigma\!\int\limits_{-\infty}^{\eta_1}B(\eta_2v+x)v d\eta_2d\eta_1\right)\right)v
 d\sigma d\tau \right|.\nonumber
\end{eqnarray}
We shall estimate each $\Delta_{2,i}$ for $i=1\ldots 6$.
From \eqref{l4.12a}, \eqref{l4.12b}, \eqref{l4.4} and \eqref{l4.5} it follows that
\begin{equation}
\max(\Delta_{2,1},\Delta_{2,2})\le {\beta_1\beta_2n^{3/2}2^{\alpha+2}(2\alpha+3)(1+\sqrt{n}|v|)\over (\alpha-1)\alpha^2(\alpha+1)
{|v|\over\sqrt{2}}({|v|\over\sqrt{2}}-R)^3(1+{|x|\over\sqrt{2}})^{2\alpha-1}}.\label{l4.13}
\end{equation}

Using \eqref{1.3}, \eqref{1.4}, \eqref{l4.3}, \eqref{14} and \eqref{l4.S1} we obtain that
\begin{eqnarray}
&&\left|B(\sigma v+x+A_{v,x}^1(0,0)(\sigma))\int_{-\infty}^\sigma (-\nabla V)(\eta v+x)d\eta\right|\nonumber\\
&&\le n^{3/2}2^{\alpha +1}\beta_1^2(1+{|x|\over\sqrt{2}}+({|v|\over \sqrt{2}}-R)|\sigma|)^{-\alpha-1}\nonumber\\
&&\times\int_{-\infty}^\sigma 
(1+{|x|\over\sqrt{2}}+{|v|\over \sqrt{2}}|\eta|)^{-\alpha-1}d\eta\label{l4.14}
\end{eqnarray}
for all $\sigma \in \R.$
From \eqref{l4.14} and \eqref{l4.12c} it follows that
\begin{equation}
\Delta_{2,3}\le {3n^{3/2}\beta_1^22^{\alpha+1}\over (\alpha-1)\alpha^2{|v|\over \sqrt{2}}({|v|\over \sqrt{2}}-R)^2(1+{|x|\over\sqrt{2}})^{2\alpha-1}}.
\label{l4.15}
\end{equation}

Using growth properties of $B$ \eqref{1.4}, \eqref{l4.3},   \eqref{14}, and \eqref{l4.S2} we obtain
\begin{eqnarray*}
&&\left|\left(B(\sigma v+x+A_{v,x}^1(0,0)(\sigma))-B(\sigma v+x)\right)\left(\int_{-\infty}^\sigma B(\eta v +x)vd\eta\right)\right|\\
&&\le n^{5/2}\beta_1\beta_22^{\alpha+2}|v|(1+{|x|\over\sqrt{2}}+({|v|\over \sqrt{2}}-R)|\sigma|)^{-\alpha-2}|A_{v,x}^1(0,0)(\sigma)|\\
&&\times\int_{-\infty}^\sigma (1+{|x|\over\sqrt{2}}+{|v|\over \sqrt{2}}|\eta|)^{-\alpha-1}d\eta,
\end{eqnarray*}
for all $\sigma \in\R$.

Using also \eqref{l4.4}, we obtain
\begin{equation}
\Delta_{2,4}\le{3n^3(2\alpha+3)\beta_1^2\beta_22^{\alpha+2}\sqrt{2}(1+\sqrt{n}|v|)\over (\alpha-1)\alpha^3(\alpha+1){|v|\over
\sqrt{2}}({|v|\over\sqrt{2}}-R)^3(1+{|x|\over\sqrt{2}})^{3\alpha-1}}.\label{l4.16}
\end{equation}

From growth property of $B$ \eqref{1.4}, and from the inequality $\max({\rho_1\over r},{\rho_2\over R})\le 1$, \eqref{l4.9},
\eqref{1.4}, \eqref{14} and \eqref{l4.S1}, it follows that
$$
\left|\left(B(\sigma v+x+A_{v,x}^1(0,0)(\sigma))-B(\sigma v+x+\int\limits_{-\infty}^\sigma\!\int\limits_{-\infty}^{\eta_1}\!\!\!
B(\eta_2v+x)v d\eta_2 
d\eta_1)\right)v\right|\\
$$
\begin{eqnarray}
&\le& n^{3/2}\beta_22^{\alpha+2}|v|(1+{|x|\over\sqrt{2}}+({|v|\over \sqrt{2}}-R)|\sigma|)^{-\alpha-2}
\left|\int\limits_{-\infty}^\sigma\!\int\limits_{-\infty}^{\eta_1}\!\!\!\nabla V(\eta_2v+x) d\eta_2 d\eta_1\right|\nonumber\\
&\le& n^2\beta_1\beta_22^{\alpha+2}|v|(1+{|x|\over\sqrt{2}}+({|v|\over \sqrt{2}}-R)|\sigma|)^{-\alpha-2}\nonumber\\
&&\times\int_{-\infty}^\sigma\int_{-\infty}^{\eta_1}(1+{|x|\over\sqrt{2}}+{|v|\over \sqrt{2}}|\eta_2|)^{-\alpha-1} d\eta_2 d\eta_1,
\label{l4.17}
\end{eqnarray} 
for all $\sigma \in \R$.
Therefore by using \eqref{l4.12e}-\eqref{l4.12f} we obtain
\begin{equation}
\Delta_{2,5}
\le {2^{\alpha+2}\sqrt{2}n^2\beta_1\beta_2\over (\alpha-1)\alpha^2(\alpha+1){|v|\over \sqrt{2}}({|v|\over \sqrt{2}}-R)^2
(1+{|x|\over\sqrt{2}})^{2\alpha-1}};
\label{l4.18}
\end{equation}
and
\begin{equation}
\Delta_{2,6}
\le {2^{\alpha+2}\sqrt{2}n^2(2\alpha+3)\beta_1\beta_2\over (\alpha-1)\alpha^2(\alpha+1)({|v|\over \sqrt{2}}-R)^3
(1+{|x|\over\sqrt{2}})^{2\alpha-1}}.
\label{l4.19}
\end{equation}
Estimate \eqref{2.14b} follows from \eqref{l4.11}, \eqref{l4.13}, \eqref{l4.15}, \eqref{l4.16}, \eqref{l4.18} and \eqref{l4.19}.
\hfill$\Box$

\section*{References}

{\parindent=-1.7cm 
\leftskip=-\parindent
\leavevmode \hbox to 1.7cm {}
\vskip -5mm
\leavevmode \hbox to 1.7cm {[Ab]\hfill}N. H. Abel,
 Aufl\"osung einer mechanischen Aufgabe,
{\it J. Reine Angew. Math.} {\bf 1}, 153-157 (1826).
 French transl.: R\'esolution d'un probl\`eme de m\'ecanique, \OE uvres compl\`etes de Niels Henrik Abel (L. Sylow, S. Lie, eds)
 vol.1, pp97-101, Gr\o ndahl, Christiana (Oslo), 1881.

\leavevmode \hbox to 1.7cm {[AFC]\hfill}M. A. Astaburuaga, C. Fernandez, V. H. Cort\'es,
The direct and inverse problem in Newtonian Scattering,
{\it Proc. Roy. Soc. Edinburgh Sect. A} {\bf 118}, 119-131 (1991).

\leavevmode \hbox to 1.7cm {[C]}A.M. Cormack, Representation of a function by its line integrals, with some radiological applications,
{\it J. Appl. Phys.} {\bf 34}, 2722-2727 (1963).

\leavevmode \hbox to 1.7cm {[GG]}I.M.  Gel'fand, M.I. Graev, Integrals over hyperplanes of basic and generalized functions,
{\it Dokl. Akad. Nauk SSSR}  {\bf 135}, 1307--1310 (Russian). Engl. transl. : Soviet Math. Dokl.  1,   1369--1372 (1960).

\leavevmode \hbox to 1.7cm {[GGG]\hfill}I.M. Gel'fand, S.G. Gindikin, M.I. Graev, Integral geometry
in affine and projective spaces, {\it Itogi Nauki i Tekhniki, Sovr. Prob. Mat.}
{\bf 16}, 53-226 (1980) (Russian).

\leavevmode \hbox to 1.7cm {[GN]\hfill}M.L. Gerver, N. S. Nadirashvili, Inverse problem of mechanics at high energies, 
{\it Comput. Seismology} {\bf 15}, 118-125 (1983) (Russian).

\leavevmode \hbox to 1.7cm {[H]\hfill}S. Helgason, The Radon transform on Euclidean spaces, 
compact two-point homogeneous spaces and Grassmann manifolds,  {\it Acta Math.}  {\bf 113},  153--180 (1965).

\leavevmode \hbox to 1.7cm {[Jo1]\hfill}A. Jollivet, On inverse scattering for the multidimensional relativistic Newton equation at 
high energies, {\it J. Math. Phys. }  {\bf 47}(6), 062902 (2006).

\leavevmode \hbox to 1.7cm {[Jo2]\hfill}A. Jollivet, On inverse scattering in electromagnetic field in classical relativistic mechanics at high
energies, 2005 preprint, 

\noindent /math-ph/0506008.

\leavevmode \hbox to 1.7cm {[Jo3]\hfill}A. Jollivet, On inverse problems in electromagnetic field in classical mechanics at fixed energy, {\it J. Geom. Anal.} {\bf 17}:(2), 275-320 
(2007).

\leavevmode \hbox to 1.7cm {[K]\hfill}J. B. Keller,
Inverse problems,
{\it Amer. Math. Monthly} {\bf 83}, 107-118 (1976).

\leavevmode \hbox to 1.7cm {[LL1]\hfill}L.D. Landau, E.M. Lifschitz, {\it Mechanics}, Pergamon Press Oxford, 1960.

\leavevmode \hbox to 1.7cm {[LL2]\hfill}L.D. Landau, E.M. Lifschitz, {\it The Classical Theory of Fields}, Pergamon Press New York, 
1971.

\leavevmode \hbox to 1.7cm {[LT]\hfill}M. Loss, B. Thaller, Scattering of particles by long-range magnetic fields, {\it Ann. Physics} {\bf 176}, 159-180 (1987).

\leavevmode \hbox to 1.7cm {[Na]\hfill}F. Natterer, {\it The Mathematics of Computerized Tomography}, Stuttgart: Teubner and  
Chichester: Wiley, 1986

\leavevmode \hbox to 1.7cm {[No]\hfill}R.G. Novikov, Small angle scattering and X-ray transform
in classical mechanics, {\it Ark. Mat.} {\bf 37},  141-169 (1999).

\leavevmode \hbox to 1.7cm {[R]\hfill}J. Radon, \" Uber die Bestimmung von Funktionen durch ihre Integralwerte l\" angs 
 gewisser Mannigfaltigkeiten. {\it Ber. Verh. S\" achs. Akad. Wiss. Leipzig, Math.-Nat. K1} 
{\bf 69}, 262-277 (1917).

\leavevmode \hbox to 1.7cm {[S]\hfill}B. Simon, Wave operators for classical particle scattering.
{\it Comm. Math. Phys.} {\bf 23}, 37-48 (1971).

}

\vskip 8mm

\noindent A. Jollivet

\noindent Laboratoire de Math\'ematiques Jean Leray (UMR 6629)

\noindent Universit\'e de Nantes 

\noindent F-44322, Nantes cedex 03, BP 92208,  France

\noindent e-mail: jollivet@math.univ-nantes.fr

\end{document}